\documentstyle[aps,epsf,amssymb,twocolumn]{revtex}
\begin{document}
\global\firstfigfalse
\global\firsttabfalse
\bibliographystyle{prsty}
\title{Drag and Hall drag in a bi--layer system with pinholes}
\author{Yuval Oreg and Bertrand I. Halperin}
\address{ Lyman Laboratory of Physics, Harvard University, Cambridge
  MA 02138 \\ \medskip\medskip \medskip{}~{ } \parbox{14cm} 
  {\rm The transresistance and the Hall transresistance of dirty
    two--dimensional bi--layer systems in the presence of tunneling
    bridges (pinholes) are studied theoretically.  We find, at weak
    magnetic field, a non--zero Hall transresistance.  In a geometry
    where the pinholes are concentrated in the middle of the sample, a
    quantum process gives the dominant contribution to both the
    ordinary transresistance and the Hall transresistance.  Arising
    from the interplay between Coulomb repulsion, disorder and
    tunneling, the quantum contribution increases in a singular way as
    the temperature decreases.\\
    \smallskip \\
%    \centerline{\Large Date: \today}
 } } \maketitle
\section{ Introduction}

\label{se:intro}

The progress in micro--structure technology of semi--conductors has
made it possible to fabricate a pair of parallel two dimensional (2D)
electronic layers that are spatially close to each other.
Experimental
\cite{DR:Solomon89,DR:Gramila91,DR:Eisenstein92,DR:Gramila93,DR:Gramila94,DR:Sivan92,DR:Giordano94}
and theoretical efforts
\cite{DR:Laikhtman90,DR:Jauho93,DR:Zheng93,DR:Kamenev95,DR:Flensberg95,DR:Ussishkin97,DR:Kim96,DR:Oreg98}
are being performed to understand the behavior of these systems.  The
physics of bi--layer system is interesting in its own right , but it
also serve as a tool to test the internal layer properties.

\begin{figure}[h]
\vglue 0cm \hspace{0.05\hsize} \epsfxsize=0.8\hsize
 \epsffile{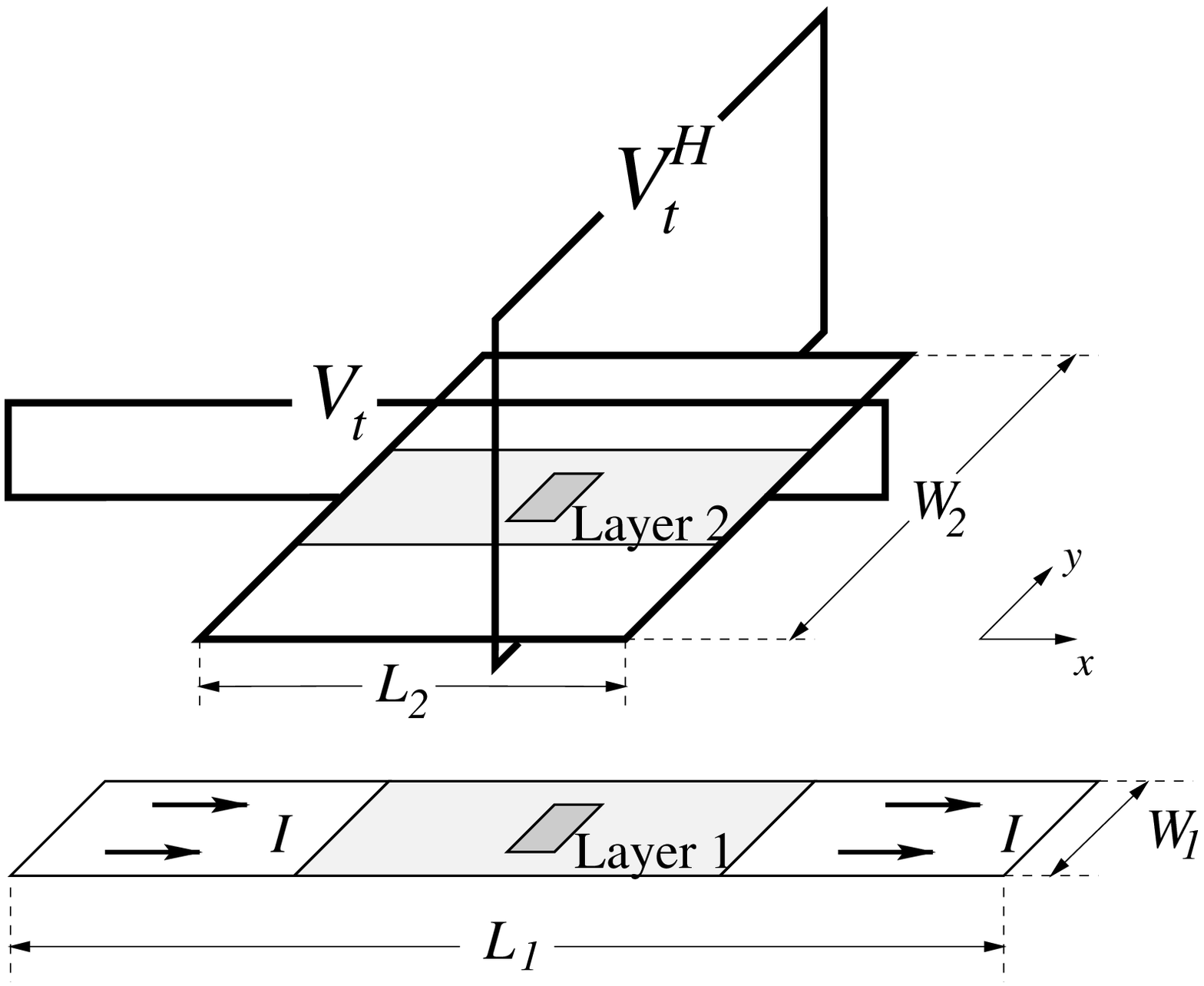} 
\refstepcounter{figure}
  \label{fg:setup} \\ 
  {\small FIG.\ \ref{fg:setup} Geometry for a drag experiment. The
    lightly shaded areas denote the overlapping regions of the 2D
    electron gases.  In a typical transresistance measurement a
    current $I$ is sent through the 2D layer 1 and a (trans)potential
    $V_t$ is developed in the 2D layer 2.  In the absence of tunneling
    between the layers, the Hall transpotential $V_t^H$ is zero at a
    weak magnetic field $\vec H$ perpendicular to the layers.  In the
    text we discuss the influence of tunneling through local pinholes
    (or bridges) between the layers on the transresistance and the
    Hall transresistance [defined in Eqs.~(\ref{def:Rt}) and
    (\ref{def:RtH})], in the presence of Coulomb repulsion.  The dark
    shaded regions denote the area where pinholes exist and tunneling
    between the layers is possible.  We consider cases where the
    tunneling region is much smaller then the overlapping region and
    where the two regions are equal.}
\end{figure}

In a typical experimental set up, a current $I$ is sent through one of
the layers (layer 1) and a voltage drop, $V_t$, is measured by
separate contacts on the other one (layer 2) (see
Fig.~\ref{fg:setup}). The ratio between $V_t$ and $I$ defines the
transresistance, $R_t$. In a similar way we can define a Hall
transresistance $R_t^H$, when a magnetic field is applied. [For
precise definitions see Eqs.~(\ref{def:Rt}) and (\ref{def:RtH}) in
Sec.~\ref{se:setup}.]
The behavior of the transresistances is rather rich due to the
possibility to control the layer areas, the electron density in the 2D
layers, their mobility, the interlayer tunneling rate, and external
parameters like the temperature and a magnetic field.

In this work we concentrate on the corner of the parameter space where
the mobility is relatively low (disorder is large), a tunneling
between the layers through local pinholes (or bridges) is possible and
a weak magnetic field can be turned on. This situation occurs when the
average distance between the layers, $d$, is relatively small so we
expect the Coulomb forces to be dominant\cite{DR:Bonsager98}. [We
consider here only the cases of a weak or zero magnetic field, with
weak interlayer tunneling and weak interlayer interaction;
generalizations to strong magnetic fields and nonperturbative
tunneling and interactions are subjects for future studies.]

In the absence of tunneling, a transresistance arises by frictional
(drag) forces due to the Coulomb repulsion between electrons in the
two layers. This mechanism involves thermal density fluctuations
around a mean value of the electron density and therefore vanishes at
low temperatures. In the remainder of the article, this effect will be
referred to as the {\em classical drag} mechanism.

In the presence of tunneling through pinholes there are additional
physical processes that lead to a finite transresistance.  The first
has to do with the fact that, in the presence of pinholes, current can
flow from layer 1 to layer 2 through a pinhole close to the source
lead and flow back through another pinhole close to the drain lead.
This purely classical effect leads to a net current flow in layer 2,
to a potential drop and finally to finite transresistance, which we
refer to as the {\em leakage contribution}.  The potential due to this
mechanism depends on the distribution of the pinholes in the sample
and, through the weak temperature dependence of the pinholes'
conductance, on the temperature.

In addition, there is a more subtle mechanism due to the interplay
between Coulomb repulsion, tunneling and disorder\cite{DR:Oreg98}. We
shall see that the main part of this effect arises from frequencies
that are larger than the temperature [see discussion after
Eq.~(\ref{eq:Fa})].  In that sense, the effect involves virtual
processes of quantum origin, and will be called the {\em quantum
  mechanism}. The quantum effect is a generalization of the intralayer
interaction correction in a single layer disorder system
\cite{DS:Altshuler79} to the geometry of bi--layer systems.

Unlike the classical drag mechanism, which vanishes at low
temperature, and the leakage contribution, which depends weakly on the
temperature, due to the electrons diffusive motion in layer 1 and
layer 2 {\em the quantum mechanism increases in a singular way when
  the temperature decreases}.

Without pinholes the Hall transresistance vanishes rapidly ($\propto
T^4$) at low temperatures \cite{DR:Kamenev95,DR:Kuang97}. (We do not
discuss the case where there are strong correlations between the
layers \cite{DR:Yang98}.) We will see below, however, that in the
presence of tunneling, there are nonvanishing leakage and quantum
contributions to the Hall transresistance. Thus, a measurement of the
Hall transresistance is a direct measurement of the leakage and
quantum effects.

In most geometries, it might be difficult experimentally to separate
the quantum and the leakage contributions because the temperature
dependence of the transresistance measurements has to be very accurate
\cite{DR:Oreg98}.  However, we argue below that in a geometry where
the pinhole distribution is deliberately concentrated near the middle
of the sample, at low enough temperature the quantum mechanism is
larger than both the classical drag mechanism and the leakage
contribution.  While the standard drag measurement in the absence of
tunneling gives information on thermal fluctuations and interlayer
interactions of the system, in the presence of tunneling through
pinholes the transresistance measurements can provide interesting
information about quantum processes that involve an interplay between
disorder and interaction.
For simplicity we quote here results for 2D layers that have identical
properties; i.e., they have the same sheet resistance $R_\square$,
diffusion constant $D$, Fermi energy $E_F$, Fermi momentum $k_F$, mean
free path $l$ and mean free time $\tau$, total density of states
(including spin) $\nu$ and inverse Thomas-Fermi screening length
$\kappa=2\pi e^2 \nu$.

Since the measured transvoltages $V_t$ and $V_t^H$ depend on the
precise location of the voltage contacts, we use here an average over
the voltages along the appropriate edges in the definition of the
transresistances; e.g., for rectangular geometry we integrate the
potential along the boundary and divide by its length. [For precise
definition of the transresistances see Sec. \ref{se:defs}.]  We assume
that the length $L_1$ of layer 1 is large compared to its width $W_1$,
so that the current density is uniform through the interaction region.
A discussion of the actual voltage distribution is given in
Sec.~\ref{se:V2} below, for several cases of interest. 

In case where $L_2 \gg W_2$ and the measuring probes for the
longitudinal transresistance $R_t$ are very far from the tunneling
places, the potential in layer 2, $V^{(2)}(x,y)$, is practically
independent of $y$ near the points of measurement.  In that case the
average over $y$ in the definition of $R_t$ [see Eq.~(\ref{def:Rt})]
is not needed.  However, in this geometry the Hall transvoltage must
be measured close to the tunneling point and is sensitive to the
distance from it along the $x$ axis.  [In fact, as shown in
Sec.~\ref{se:middle}, the Hall transvoltage falls off exponentially
with the distance from the tunneling points.] To avoid this factor, it
is preferable to measure the Hall transvoltage in a ``cross'' geometry
where $L_2 \approx W_1 \ll W_2$. In that case the Hall transresistance
is measured far from the tunneling points, the potential
$V^{(2)}(x,y)$ depends weakly on $x$, and the average over $x$ in
Eq.~(\ref{def:RtH}) is not needed.

We shall see, below, that $R_t$ and $R_t^H$ may be written as the $x$
and $y$ components of a vector $\vec R_t$, which has the form
\begin{equation}
\label{eq:Rtint}
\vec R_t = \frac{1}{I W_2}  \left[ {\vec P} + {\vec F} \right],
\end{equation}
where $I$ is the total current flowing in layer 1, and $W_2$ is the
width of layer 2 (see a generalization for nonrectangular
configurations in Sec.\ref{se:defs}).  The components of the vector
$\vec P$, arising from the leakage contribution, are given by the
product of resistivity tensor of layer 2 and the dipole moment of the
tunneling current distribution. The vector $\vec F$ is due to the
momentum transferred from layer 1 to layer 2, it includes both the
classical drag contribution and the quantum effect in the presence of
tunneling between the layers.

If the tunneling between the layers occurs uniformly all over the
sample then the dipole moment $\vec P$ is large and the quantum
mechanism is a small correction to the leakage contribution to the
transresistances. However, when the pinholes are concentrated in the
middle of the samples the dipole moment is small and the quantum
contribution is dominant.

When layer 1 and layer 2 have rectangular shapes of sizes $L_1 \times
W_1$ and $L_2 \times W_2$ respectively and the pinholes distribution
is concentrated in a rectangle of dimensions $a \times b$ near the
middle of the sample we find:
\begin{eqnarray}
R_t= - \frac{S_{\rm int}}{W_1 W_2} \rho_D +\left(\frac{ a^2}{W_1
W_2}\left[1+  \alpha_t t_\square \log\left(
\frac{1}{T \tau} \right) \right] \right. \nonumber \\
\label{eq:Rt}
\left.
 \;\;\;\;\;\;\;\;\;\;\;\;\;\;\;\;+ t_\square \pi
\frac{\log(\kappa d)}{ \kappa d} \frac{L_T^2}{W_1 W_2} \right)
\frac{R_\square^2}{12 R_\perp},
\end{eqnarray}
and
\begin{eqnarray}
R_t^H= \left(\frac{ a^2+b^2}{2 W_1 L_2}\left[1 + \alpha_t^H 
t_\square\log\left(\frac{1}{ T \tau} \right) \right] \right. \nonumber \\
\label{eq:RtH}
\left. \;\;\;\;\;\;\;\;\;\;\;\;\;\;\;\; + t_\square \pi \frac{\log(\kappa d)}{ \kappa d} \frac{L_T^2}{W_1 L_2} \right)
 \frac{ R_\square R_H}{6 R_\perp}.
\end{eqnarray}
where $\rho_D \propto T^2 $ is the bulk drag coefficient [see the
precise expression for it in Eq.~(\ref{eq:rhoD})], $t_\square =
R_\square e^2 / 2 \pi^2 \hbar$, $S_{\rm int}= L \times W$ is the area
of the layers' overlapping region, $L=\min \{L_1,L_2\}$ and $W=\min \{
W_1, W_2 \}$, $R_H$ is the Hall resistivity of a single layer, and
$R_\perp$ is the total resistance between the layers. The term
proportional to $a^2$ (or $a^2+b^2$ for the Hall transresistance) is
from the leakage contribution and the term proportional to $L_T^2 =
D/T$ is the quantum contribution.  The coefficients $\alpha_t$ and
$\alpha_t^H$ are numbers of order unity and the corresponding terms
are due to the zero bias anomaly correction to $R_\perp$, and due to
intralayer interaction and weak localization corrections to the
conductivity within each layer \cite{RFS:Altshuler85} (see also
Sec.~\ref{sse:rev}).

Expressions (\ref{eq:Rt}) and (\ref{eq:RtH}) are valid for 
\begin{equation}
\label{eq:validity}
T > \max \{ \frac{D}{L^2_{\min}}, \frac{1}{\tau} e^{-\pi
R_\perp/R_\square} \}
\mbox{ and } H \le \frac{1}{\mu},
\end{equation}
where $L_{\min} = \min \{L,W \}$, $H$ is the external magnetic field
perpendicular to the layers, and $\mu$ is the sample mobility.

If $L_2$, $W_2$ and $W_1$ are all comparable to each other, then the
quantum contribution simply flattens out and becomes temperature
independent for $T< D/L_{\min}^2$. On the other hand if $L_2$ is much
larger than $W_1$ and $W_2$, then there could be an intermediate
temperature $D/L_2^2 \ll T \ll D/L_{\rm min}^2$ where the system is
quasi one--dimensional, and the temperature dependence of the quantum
contribution to $R_t$ may be even more singular then in
Eq.~(\ref{eq:Rt}).  If $R_\perp$ is not very large so that the energy
scale $(1/\tau) \exp\left[-\pi R_\perp/R_\square\right]$ may be larger
than $D/L^2_{\min}$, then effects which are non linear in
$R_\perp^{-1}$ may need to be taken into account at low temperatures.
If $H$ exceeds $1/\mu$ then effects nonlinear in magnetic field become
important.

We have also assumed through out the paper that the current in layer 1
is so low that the cut off of the quantum process is determined by the
temperature and not by the voltage difference between layer 1 and
layer 2. This assumption should hold as long as $J R_\square L_T, J
R_\square b \ll T/e$ where $J$ is the current density in layer 1.

Let us examine now expressions (\ref{eq:Rt}) and (\ref{eq:RtH}) for
the transresistances. If we further assume that $a,b \ll L_T \ll
L_{\min}$ then the leakage contribution is suppressed and the quantum
contribution is larger than the leakage contribution.  On the other
hand, the classical drag contribution could be larger than the quantum
contribution at finite temperatures, even though $\rho_D$ vanishes as
$T^2$, because the classical drag is effective over the entire area of
the overlap region, as reflected in the prefactor $S_{\rm int}$.  In
order to minimize the classical drag contribution, one should choose the
dimension of the sample as small as possible, consistent with the
requirement that $L_{\min}$ remain larger than $L_T$ at the lowest
achievable temperatures.

By contrast, there is no contribution from the classical Coulomb drag
to the Hall transvoltage. This happens because in the absence of
tunneling no current is flowing in layer 2, there is no Lorentz force
that should be compensated, and as a result no Hall transvoltage is
developed at low temperatures \cite{DR:Kuang97}.

In case when $a,b$ are comparable to the layer size, i.e., when the
pinhole distribution is uniform, the quantum contribution is a small
correction to the leakage contribution, similar to the small
interaction corrections in single layer substances
\cite{RFS:Altshuler85}. In that case the temperature dependence is
mainly determined by the intralayer interaction, and weak localization
corrections to $R_\square$ and the zero bias anomaly correction to
$R_\perp$.

Examining Eq.~(\ref{eq:Rt}) we see that at low temperatures the
contribution from the classical drag vanishes and the contribution
from the leakage term $\propto a^2$ is a temperature--independent
constant. The quantum contribution increases as $1/T$. Thus with the
right choice of parameters and at low enough temperatures the quantum
contribution is dominant. We note that while the usual drag mechanism
leads to a potential drop in layer 2 that is opposite to the voltage
drop in layer 1, both the leakage and the quantum mechanisms give rise
to a potential drop in the same direction as in layer 1.  Therefore
$\rho_D$ has a sign which is opposite to leakage and the quantum
contributions. Thus we expect $\cite{DR:Raichev97}$ that at a
temperature $T^*$ the transresistance will change signs. The Hall
transresistance has no contributions from the classical drag mechanism
\cite{DR:Kamenev95}, hence, it gives a direct measurement of the
leakage and the quantum contribution.

For a $GaAs/ AlGaAs$ rectangular sample of dimensions $20 \mu m \times
5\mu m$, mobility $\mu= 5 \times 10^4 {cm^2}/{Vs}$, electron density
$n=4 \times 10^{10} cm^{-2}$, $R_\perp=20 k\Omega$ and $a=0.1\mu m,
b=0.1\mu m$, yielding $R_\square \cong 3 k\Omega$, we find (with
$\kappa d \cong 3$) that the total contributions of the classical,
leakage and quantum mechanisms (neglecting zero bias anomaly, weak
localization and intralayer interaction corrections) are:
\begin{equation}
\label{eq:Rtest}
R_t (m \Omega) \cong -300 T^2  + 16 + \frac{4}{T},
\end{equation}
with $T$ measured in Kelvin, for $T > 2 mK$.  At $T^* \cong 0.3 K $
the transresistance is zero. This means that the leakage and quantum
contributions win over the classical drag and at $T < 0.2 K$ the
quantum contribution is larger than the leakage one. At $T\cong 2 mK$
the system become quasi--$1D$ and the behavior of the quantum
corrections is even more singular $\propto 1/ T^{3/2}$ [See
Eq.~(\ref{eq:Q1D})], eventually the at $T \cong 0.1mK$ the quantum
contribution becomes temperature independent.

The Hall transresistance in the ``cross'' geometry, i.e., when the
dimensions of layer 1 are $20 \mu m \times 5\mu m$ and of layer 2 are
$5 \mu m \times 20 \mu m$ with the same 2D electron gas parameters as
above, is found to be
\begin{equation}
\label{eq:RtHest}
R_t^H (m \Omega) \cong (160 + \frac{40}{T}) H,
\end{equation}
with $H$ measured in Tesla, for $T> 2mK$, $H< 0.2 {\rm T}$.  The last
condition insures that we are in the linear regime with respect to the
magnetic field. Since in the absence of tunneling the Hall
transresistance is zero, this finite Hall transresistance is a direct
measurement of the leakage and quantum corrections.

We note that the tunneling region in the middle of the sample does not
have to be a square. If it has the shape of a slit geometry, e.g.,
$a=0.01 \mu m$ and $b=1 \mu m$ then the leakage term in $R_t$ is even
smaller. One should have in mind, though, that in this case the slit
has to be aligned very precisely perpendicular to the current in layer
1, in order to keep the leakage term small. In the slit geometry it is
easier to make $R_\perp$ larger.

The structure of the remainder of the paper is as follows: in
Sec.~\ref{se:setup} we discuss the formulation of the problem in terms
of the resistivity tensor $\rho_{ij}^{\alpha \beta}({\vec r}, {\vec
  r'})$, the appropriate boundary conditions and the continuity
equation. This leads to a set of integro--differential equations
(\ref{eq:con}), (\ref{eq:ohmslawf}) and (\ref{eq:bc12}). Their
solution, combined with the appropriate generalization of Gauss's law,
gives the transresistances in terms of the conductivity tensor
$\sigma_{ij}^{\alpha \beta}({\vec r}, {\vec r'})$, inverse of the
resistivity tensor, which can be determined using a Kubo formalism.
In Sec.~\ref{se:micro} a microscopical analysis of different parts of
$\sigma_{ij}^{\alpha \beta}({\vec r}, {\vec r'})$ is performed using
the linear response formalism (which we generalized to include the
tunneling through local pinholes) for dirty materials.  Later we
discuss in some more details the potential distribution in
\mbox{layer~2.}  In Sec.~\ref{se:middle} we solve the
integro--differential equations in the case where the tunneling occurs
only in the middle.  We find that the current flow lines in layer 2
are similar to the field lines of a dipole in 2D.  In
Sec.~\ref{se:uni} we discuss the case when the pinhole distribution is
uniform. We solve Eq.~ (\ref{eq:con}), (\ref{eq:ohmslawf}) and
(\ref{eq:bc12}) perturbatively and find expression for the potential
in \mbox{layer 2.} Finally, after a concluding section, appendixes
with details of calculations are presented.

\section{Macroscopic Equations}

\label{se:setup}
To be specific we will analyze a system with the geometry depicted in
Fig.~\ref{fg:setup}.  We use a notation where the indices $i,j,k$ run
over the directions $x,y$ and $\alpha,\beta=1,2$ are layer indices.
Summation over repeated indices is understood. The local current (per
unit length) in layer $\alpha$ and direction $i$, $J_{i}^{\alpha}$, is
related to the potential difference between the layers by the
continuity equation:
\begin{equation}
\label{eq:con}
\nabla_i J_i^\alpha({\vec r})=(-1)^\alpha J_T\left({\vec
r}\right)= g^t({\vec r}) A^{\alpha \beta}V^{\beta}({\vec r}),\;\;\;
\end{equation}
where $J_T\left({\vec r}\right)$ is the tunneling current density
between the layers, the matrix $A^{\alpha \beta}$ is $-1$ if $ \alpha =
\beta $ and $ 1 $ if $ \alpha \ne \beta$,
$$
%\label{def:gt}
g^t({\vec r}) = \sum_l \delta ({\vec r} - {\vec R}^l)  g^l, \;\;\;\;
g^l= \frac{e^2}{2 \pi \hbar} |t^l|^2,
$$
and $t^l$ is the transmission amplitude of the $l$th pinhole located
at ${\vec R}^l$.
In addition, the current at point ${\vec r}$ is related to the
electric fields via a generalized Ohm's law that includes the momentum
transfer from the other layer:
\begin{equation}
\label{eq:ohmslawf}
J_i^\alpha({\vec r}) = \sigma^{(\alpha)}_{ik} \left[ -\bbox{\nabla}_k V^{(\alpha)}
({\vec r}) + f^{(\alpha)}_k({\vec r}) \right],
\end{equation}
where $\sigma_{ij}^{(\alpha)}$ are the conductivities of layer
$\alpha$ in the absence of the other layer and the vector $\vec
f^{(\alpha)}$, describing momentum transfer from layer $\beta \ne
\alpha$ to layer $\alpha$, is defined below.  To complete the set of
the differential equations (\ref{eq:con}), (\ref{eq:ohmslawf}) we
impose the following boundary conditions. (i) No current can flow
perpendicular to lateral edges of layer 1 or to the boundaries of
layer 2; (ii) the current enters and leaves layer $1$ with a uniform
current density $J$.  When layer 1 has the shape of a rectangular of
length $L_1$ and width $W_1$, the boundary conditions on
$J^\alpha_i({\vec r})$ are
\begin{mathletters}
\label{eq:bc12}
\begin{equation}
\label{eq:bc1}
J^{(1)}_{x}(\pm L_1/2,y)=J, \;\;\;\; J^{(1)}_{y}(x,\pm W_1/2)=0,
\end{equation}
and 
\begin{equation}
\label{eq:bc2}
\left. n_i J_i^{(2)} \right|_{\partial S_2} = 0,
\end{equation}
\end{mathletters}
where $S_2$ is the region of layer $2$, $\partial S_2$ its boundary
and $\vec n$ is a vector normal to $\partial S_2$. Notice that, since
charge can not be accumulated in layer 2, (\ref{eq:bc2}) is possible
only if \mbox{$\int\!\!\!\int_{S_2} d^2 r \nabla \cdot \vec J^{(2)}=0$}.
A solution of (\ref{eq:con})-(\ref{eq:bc12}) gives the
transresistances in terms of $\vec f^{(\alpha)}$.  [Different boundary
conditions reflecting different experimental configurations can be
analyzed in a way similar to the one discussed below, they will change
the results for the measured transvoltages.]

The components of the vector $\vec f^{(\alpha)}$ are given by 
\begin{equation}
\label{def:f}
 f^{(\alpha)}_k({\vec r}) \equiv \int\!\!\!\!\int_{S_\beta} d^2  r' \tilde
\rho^{\alpha \beta}_{kj}({\vec r}, {\vec r'}) J_j^{\beta}({\vec r'}),
\end{equation}
where
\begin{equation}
\label{def:tilderho}
 \tilde \rho^{\alpha \beta}_{kj}({\vec r}, {\vec r'}) \equiv
\rho^{(\alpha)}_{kj}
\delta_{\alpha \beta} \delta({\vec r}-{\vec r'})
- \rho^{\alpha \beta}_{kj}({\vec r}, {\vec r'}).
\end{equation}
The symbol $\rho^{\alpha \beta}_{kj}({\vec r}, {\vec r'})$ is a matrix
in the variables $\vec r$ and $ \vec r'$, and in the layer and
Cartesian indices, which is the matrix--inverse of the conductivity
tensor $\sigma_{i j}^{\alpha \beta} ({\vec r}, {\vec r'})$ that can be
found from a microscopic treatment (see Sec.~\ref{se:micro})]. The
tensor $\rho^{(\alpha)}_{kj} \delta({\vec r}-{\vec r'}) \equiv
\rho^{\alpha \alpha}_{kj}({\vec r}, {\vec r'})$ is the resistivity
tensor of layer $\alpha$ in the absence of the other layer (inverse to
$\hat \sigma^{(\alpha)}$), and $S_\beta$ denotes the region of layer
$\beta$. The electric field $f^{(\alpha)}_k({\vec r)}$ describes the
field formed in layer $\alpha$ due to processes that involve the other
layer. For the case of weak coupling between the layers, which we
consider here, only the elements of $\tilde \rho$ which are
off--diagonal in the layer index need be included, as the diagonal
elements are negligible.

There are two essential contributions to $\tilde \rho$: one is due to
frictional forces (the standard classical drag) and the other is
related to the quantum mechanism mentioned earlier. [The leakage
contribution is captured by the continuity equation~(\ref{eq:con}).]
The drag mechanism does not involve any tunneling through local
pinholes between the layers.  In this mechanism the electrons in layer
2 interact with thermal fluctuations of the electrons current density
in layer 1 via Coulomb forces.  They rectify them and in this way are
dragged in the direction of the current in layer 1.  This process
leads to accumulation of charges at the edges of layer 2. As in the
case of a standard Hall effect, the charge accumulated at the edges
develops an electric field that opposes and cancels the force on the
electrons due to drag.  Therefore the voltage drop in layer 2 is
opposite to the direction of the current in layer 1, i.e., the drag
contribution, $\rho_D$, to the transresistance, $R_t$, has a negative
sign.  Since the amount of fluctuation is proportional to the
temperature, $T$, and due to the exclusion principle the average
energy of the particle hole excitations participating in the
rectification effect is proportional to $T$ as well, $\rho_D \propto
T^2$\cite{DR:Zheng93}, i.e., it vanishes as $T \rightarrow 0$.

The second mechanism contributing to $\tilde \rho$ is the quantum
process.  In this process, an electron--hole pair created in one of
the two layers, tunnels into the second layer and is annihilated
there.  The creation and annihilation processes occur due to Coulomb
interactions with the other electrons in the system, which do not
tunnel between the layers, but which take part in a density
fluctuation that is shared between the two layers as a result of the
interlayer Coulomb interaction.  As we shall see in Section
\ref{sse:calQ} below, this process gives rise to correlations in the
momenta of the two layers, in the absence of an applied electric
field, and hence a mechanism for momentum transfer when a field is
applied to one of the layers.

To solve Eqs.~(\ref{eq:con})-(\ref{eq:bc12}) perturbatively we  define:
\begin{eqnarray}
\label{def:V1}
V^{\alpha}(\vec r)& =& V^{\alpha}_{(0)}({\vec r})+V^{\alpha}_{(1)}({\vec r}), \nonumber \\
\vec J^{\alpha}(\vec r)& =&\vec J^{\alpha}_{(0)}({\vec r})+\vec J^{\alpha}_{(1)}({\vec r})
\end{eqnarray}
where $V_{(0)}^{\alpha}({\vec r}) \; \left(\vec
J_{(0)}^{\alpha}({\vec r})\right)$ is the potential (current) in the
absence of coupling between the layers and $V_{(1)}^{\alpha}({\vec
r})\; \left(\vec J_{(1)}^{\alpha}({\vec r})\right)$ is
proportional to the first order of coupling between the layers.
Lack of coupling to the second layer reduces Eq.~(\ref{eq:ohmslawf})
to
\begin{equation}
\label{eq:J0}
{J_i}_{(0)}^\alpha ({\vec r}) =
 -\sigma^{(\alpha)}_{ik} \nabla_k V^{(\alpha)}_{(0)}(\vec r).
\end{equation}
 Substitution of Eq.~ (\ref{eq:J0})in Eq.~(\ref{eq:con}) (with
vanishing right hand side for the discussed case) and using Onsager
relations [see details in Eq.~(\ref{eq:conten})] we find the Laplace
equation $\Delta V_{(0)}^\alpha=0$ with the boundary conditions
(\ref{eq:bc12}). The solution is straight forward and given by
\begin{equation}
\label{sol:V0}
V_{(0)}^{(1)}= - J \rho^{(1)}_{xx} x - J
\rho^{(1)}_{xy} y,\;\;\; V_{(0)}^{(2)}= \mbox{ const }
\end{equation}
\begin{equation}
\label{sol:J0}
{J_x}_{(0)}^{(1)}= J,\; {J_y}_{(0)}^{(1)}= 0;  \;\;\; 
{J_x}_{(0)}^{(2)}={J_y}_{(0)}^{(2)}=0.
\end{equation}
Treating the coupling between the layers perturbatively, the current and the 
potential in layer 1 are unaffected, to first order, while 
\begin{equation}
\label{eq:J2}
{J_i}_{(1)}^{(2)}({\vec r}) = \sigma^{(2)}_{ik} \left[ -\nabla_k V^{(2)}_{(1)}({\vec
r})+  f_k^{(2)} ({\vec r}) \right]
\end{equation}
where
\begin{equation}
\label{eq:f2}
f^{(2)}_k ({\vec r}) = \int d^2 r' \tilde \rho_{kj}(\vec r, \vec r')
{J_j}_{(0)}^{(1)}(\vec r').
\end{equation}
The potential pattern in the second layer depends on the location of
the pinholes, and the measured resistances generally depend on the
locations of the voltage contacts. Therefore we define integrated
potentials by:
\begin{equation}
\label{def:U}
U_i = -\oint_{\partial S_2} V^{(2)}(x,y) n_i dl = - \int \!\!\!\! \int_{S_2}
\nabla_i V^{(2)}(\vec r) d^2 r.
\end{equation}
The last equality in~(\ref{def:U}) follows from Gauss's theorem
$\int\!\!\! \int_{S_2} \nabla \cdot \vec G dx dy = \oint_{\partial
  S_2} \vec G \cdot \vec n d l$ with $\vec G= \vec w V^{(2)} $ where $\vec
w$ is an arbitrary constant non zero vector. 

If $S_2$ has a shape of a rectangular of length $L_2$ and width $ W_2$
then $U_x= \int dy \left[ V^{(2)}(-L_2/2,y) - V^{(2)}(L_2/2,y)\right]$
and $U_y = \int dx \left[V^{(2)}(x,-W_2/2) - V^{(2)}(x,W_2/2)\right]$.
Thus, the $x$ and $y$ components of $\vec U$ are a generalization of
the integrated transvoltage and Hall transvoltage respectively.

Using (\ref{eq:J2}) we find, in a matrix notation:
\begin{equation}
\label{eq:U1}
\vec U = \int\!\!\!\!\int_{S_2} d^2 r \hat \rho^{(2)} \vec
J^{(2)}_{(1)}(\vec r)  - \int \!\!\!\! \int_{S_2} d^2 r \vec f^{(2)}(\vec r).
\end{equation}
In order to continue further with the first term we use the identity
$J_i^{(2)}=\nabla_j (r_i J_j^{(2)})- r_i \nabla_j J^{(2)}_j$ and
Gauss's theorem with $G_j=r_i J^{(2)}_j$. Since $\hat \rho^{(2)}$ does
not depend on space, we find
$$
\int\!\!\!\!\int_{S_2} d^2 r \hat \rho^{(2)} \vec J^{(2)}\left({\vec r}\right)
= 
$$
$$
\hat \rho^{(2)} \int_{\partial S_2} \vec r \left( \vec n \cdot \vec
  J^{(2)} \right) dl - \hat \rho^{(2)} \int\!\!\!\! \int_{S_2} d^2 r
\vec r \vec \nabla\cdot J^{(2)}_{(1)}(\vec r).
$$
The first term vanishes due to the boundary condition (\ref{eq:bc2})
and therefore we find:
\begin{equation}
\label{eq:U}
\vec U= \vec P + \vec F,
\end{equation}
where using the Eq.~(\ref{eq:con}) we have:
\begin{equation}
\label{def:P}
\vec P = -\hat \rho^{(2)} \int \!\!\!\! \int_{S_2} d^2 r \vec r J_T(\vec r)
= - \hat \rho^{(2)} \int\!\!\!\! \int_{S_2} d^2 r \vec r g^t
V^{(1)}_{(0)}\left({\vec r}\right),
\end{equation}
and the vector $\vec F$ is given by
\begin{equation}
\label{def:F}
\vec F = -\int \!\!\!\! \int_{S_2} \hat {\tilde \rho}(\vec r, \vec r') \vec J^{(1)} (\vec r')
d^2r d^2 r'.
\end{equation}

To summarize, we find in this section that the $x$ and $y$ components
of $\vec U$ are related to the integrated transvoltage and Hall
transvoltage and have two contributions.  The first term, $ \vec P$
describes the leakage contribution to the transvoltages. The second
term $\vec F$ describes the momentum transfer from the first layer to
the second due to the classical drag effect and due to the quantum
contribution that exist only in the presence of tunneling.

We have not actually assumed that the magnetic field is weak in this
section.  However, when the Hall angle becomes very large, the
assumption that the current density is uniform in layer 1 may become
inappropriate for a fixed experimental geometry. In particular, if the
length $L$, is not much larger then $W$, current maybe concentrated
near the the edges of the sample, if the Hall angle is very large.

 \section{Definitions and calculation of the transresistance and the
 Hall transresistance}
\label{se:transresistances}

   \subsection{$\vec F$: Momentum transfer due to frictional forces
and quantum effects}
\label{se:F}

In the solution of Eqs.~(\ref{eq:con})-(\ref{eq:bc12}) given in
Eqs.~(\ref{def:U})-(\ref{def:F}) we have assumed that the couplings to
the second layer (due to frictional forces and/or tunneling) are weak
and treat them perturbatively.  In that approximation $\tilde \rho$
has only off diagonal elements (in the layer index) given by
\begin{equation}
\label{eq:rhoapp}
\tilde \rho^{\alpha \beta}_{kj}({\vec r}, {\vec r'}) \approx
 \rho_{kl}^{(\alpha)}  \sigma^{(\alpha \beta)}_{li}({\vec r}, {\vec r'}) 
\rho^{(\beta)}_{ij}\left(1-\delta_{(\alpha \beta)} \right).
\end{equation}
Notice that in the last equation the layer indices are in parentheses
to emphasize that here there is no summation over these repeated
indices.

For simplicity we assume henceforth that the layers are identical and
isotropic; extension to non identical layers is straight forward.  We
may then use the following notation for different parts of the
conductance tensor $\sigma_{ij}^{\alpha \beta}({\vec r}, {\vec r'})$:
\begin{mathletters}
\label{eq:conten}
\begin{equation}
\label{eq:contena}
\sigma_{ij}^{\alpha \beta} ({\vec r}, {\vec r'})=
 \sigma_{ij} \delta_{\alpha \beta}+ 
\tilde \sigma_{ij} \left(1-\delta_{\alpha \beta} \right),
\end{equation}
where the intralayer conductivity tensor maybe approximated as local and
independent of position:
\begin{equation}
\label{eq:contenb}
\hat \sigma= \pmatrix{ \sigma & \sigma_H \cr -\sigma_H & \sigma }
\,\delta ({\vec r} - {\vec r'}),\;\;
\end{equation}
and
\begin{equation}
\label{eq:contenc}
 {\tilde \sigma}_{ij}(\vec r, \vec r') \equiv 
\sigma_{ij}^{12}( \vec r, \vec r') = 
\sigma_{ij}^{21}( \vec r, \vec r')
\end{equation}
\end{mathletters}
The conductance $\sigma$ is given by the inverse of the (bare) sheet
resistance $R_\square= 1/e^2 \nu D$. The Hall conductance coefficient
$\sigma_H$ is essentially given by $ R_H /R_\square^2$, where
$R_H=H/ne $, $n$ is the carrier density, $e$ is the carrier charge,
and $H$ is the magnetic field. There are corrections to $\sigma$ and
$\sigma_H$ due to weak localization and Coulomb repulsion in
combination with disorder \cite{RFS:Altshuler85} that should be
included. The tunneling conductance $g^l$ has, similar to $\sigma$,
corrections due to the interplay between Coulomb repulsion and
disorder, (see details in Sec.~\ref{se:micro}). [The correction to
$\sigma$ and $\sigma_H$ due to drag and interlayer tunneling are small
and will be ignored.]

We have to discuss now the behavior of $\tilde \sigma_{ij} \left({\vec
    r}, {\vec r'}\right)$. In the approximations we use, both
frictional forces and tunneling amplitudes are small and we obtain to
first order in the tunneling rate and frictional forces:
\begin{eqnarray}
\label{def:tilsigma}
\tilde \sigma_{ij}({\vec r}, {\vec r'})&=& \sum_l {\cal Q}^l_{ij}
({\vec r} - {\vec R}^l, {\vec r'} - {\vec R}^l) + \sigma^D_{ij} \delta
( {\vec r} - {\vec r'}) \nonumber \\ &&+\; O[(\sigma^D/\sigma)^2,
|t^l|^4, |t^l|^2 (\sigma^D/\sigma)].
\end{eqnarray}
The coefficient $\sigma^D_{ij}$ is related to the drag coefficient
$\rho_D$ in clean systems ($d \ll l$) \cite{DR:Zheng93},
\cite{DR:Kamenev95} by:
\begin{eqnarray}
\label{eq:sigmaD}
\sigma^D_{xx}&\equiv&\sigma^D \cong \rho_D/R_\square^2,  \nonumber \\
\sigma^D_{xy}&\equiv&\sigma^D_H= 2\sigma_H \sigma_D/ \sigma \cong 2 R_H
\rho_D / R_\square^3
\end{eqnarray}
where
\begin{equation}
\label{eq:rhoD}
\rho_D = \frac{\hbar}{e^2} \frac{\zeta(3) \pi^2}{16}
 \frac{1}{(\kappa d)^2 (k_F d)^2} \left(\frac{T}{E_F}\right)^2.
\end{equation}
[In Eq.~(\ref{eq:sigmaD}) we have neglected weak localization
corrections. If $d > l$ then $ \zeta[3] / 16 (k_F d)^2 \rightarrow
\log[D \kappa /2 d T] / 12 (k_F l)^2$. When correlations between the
disorder in the layers are included, $\rho_D$ may be multiplied by an
$O(1)$ factor \cite{DR:Gornyi98}].  

The term ${\cal Q}$ in Eq.~(\ref{def:tilsigma}) arises due to the
interplay between the interlayer tunneling and the Coulomb
interaction. We will see that the essential contributions to this term
are from frequencies larger than the temperature, i.e., it is of
quantum origin, (therefore we use the letter ${\cal Q}$ to denote it).
We find in Sec.~\ref{se:micro} that ${\cal Q}_{jk}^{l}({\vec r}, {\vec
  r'})$ has a range $\propto L_T=\sqrt{D/T}$ which increases for $T
\rightarrow 0$.  In the expression for $\vec F$ we are interested only
in the integral of $\tilde \sigma (\vec r, \vec r')$ and the diagonal
part of the matrix
\begin{equation}
\label{eq:Qij0}
{\sf Q}_{ij} =\sum_l{\sf Q}^l_{ij}= \frac{1}{L_T^2} \sum_l \int d
{\vec s} d{\vec s'} {\cal Q}^l_{ij}({\vec s}, {\vec s'}).
\end{equation}
was calculated in Ref.~\cite{DR:Oreg98} (see also Sec.~\ref{se:Qii})
and is given (to first order in tunneling) by:
\begin{equation}
\label{eq:rQl}
{\sf Q}\equiv{\sf Q}_{xx} = \frac{e^2}{ \hbar} \frac{\log(\kappa
d)}{\kappa d} \frac{1}{24 \pi} \frac{R_\square}{R_\perp}.
\end{equation}
(Notice that there is a difference of the sample area factor divided
by $L_T^2$ between the expression here and the expression in
Ref.~\cite{DR:Oreg98}. (We introduce it to make the temperature
dependence clearer.) This expression is valid as long as higher order
corrections are small and $L_T \ll L_{\min}$.

In Sec.~\ref{se:Qxy} in Eq.~\ref{eq:sQH} we show that for a weak
magnetic field
\begin{equation}
\label{eq:QH}
{\sf Q}_{H}\equiv{\sf Q}_{xy}=-{\sf Q}_{yx}= 0.
\end{equation}
Using now the definition of ${\vec F}$ in~(\ref{def:F}) with
(\ref{sol:J0}) we find to first order in tunneling and
frictional forces at weak magnetic field
\begin{equation}
\label{eq:F}
\begin{array}{ccccc}
F_x &=&  -J S_{\rm int} \rho_D &+&  J R_\square^2 {\sf Q} L_T^2 \\
F_y &=&            & & 2 J R_\square R_H {\sf Q} L_T^2, 
\end{array}
\end{equation}
where $S_{\rm int}= S_1 \cap S_2$ is the overlapping region of the
region of layer 1, $S_1$, and the region of layer 2, $S_2$.  Notice
that in the absence of tunneling $F_y=0$. The quantum corrections
$\propto {\sf Q}$ to $F_x$ and $F_y$ are related by a factor $2
R_\square/ R_H$; this is the analog of the ``rule of two'' corrections
to the Hall resistance in a single layer samples
\cite{RFS:Altshuler85}.

  \subsection{$\vec P$: The leakage contribution}
\label{se:P}
The leakage contribution to transvoltages depends crucially on the
distribution of the pinholes. When the pinholes are distributed
uniformly inside a rectangle of size $a \times b$ centered at the
origin we find
$$
{\vec P} = \frac{J}{R_\perp a b} \int_{-a/2}^{a/2} dx
\int_{-b/2}^{b/2} dy \hat \rho \vec r \left
( x/ \sigma + y \sigma_H/\sigma^2   \right) \Rightarrow
$$
\begin{equation}
\label{eq:P}
{\vec P} = \frac{ J}{12 R_\perp} \left[ \frac{a^2}{ \sigma^2} \hat x
+ \left(a^2 + b^2 \right) \frac{\sigma_H}{\sigma^3} \hat y \right].
\end{equation}
[Notice that to fulfill the requirement $\int \!\!\! \int_{S_2} d^2 r
\nabla \cdot \vec J^{(2)} =0$, we have for a symmetric distribution of
pinholes around the origin, $V^{(2)}_{(0)} = {\rm const} = 0 $ in
Eq.~(\ref{sol:V0}).]

The quantum contribution depends on the temperature.  Thus, to make
the analysis complete we should include also in the leakage
contribution temperature dependent corrections to $R_\perp$, $\sigma$
and $\sigma_H$ arising from intralayer electron--electron interactions
and weak localization corrections (see Sec.~\ref{sse:rev}). Including
these corrections we find:

\begin{equation}
\label{eq:Pc}
\begin{array}{ccc}
{\vec P}&=& \frac{ J a^2 R_\square^2}{12 R_\perp} \left[1+ \alpha_t t_\square 
\log\frac{ 1}{T \tau } \right]
\hat x \\ &&+ \frac{ J \left(a^2 + b^2 \right) R_H R_\square}{12
R_\perp} \left[1 + \alpha_t^H
t_\square\log\frac{1} {T \tau} \right] \hat y ,
\end{array}
\end{equation}
where $\alpha_t = 2 \alpha_{\rm wl}+ 2 \alpha_{\rm in} - \beta_{\rm
  zba}$, $\alpha_t^H= 3 \alpha_{\rm in} + \alpha_{\rm wl} - \beta_{\rm
  zba} $ and $t_\square = R_\square e^2/ 2 \pi^2 \hbar$. The
coefficients $\alpha_{\rm wl}, \alpha_{\rm in}, \beta_{\rm zba}$
defined in Eqs.~(\ref{eq:sig}) and (\ref{eq:tl}) are associated with
weak localization, intralayer Coulomb repulsion, and Coulomb blocking
(Zero bias anomaly) of the tunneling between the layers. In zero
magnetic field the weak localization corrections are cut off by the
dephasing time, $\tau_\phi$ (that is proportional to $1/T$ in 2D).  In
case where $H \gg \hbar c /D \tau_\phi$ we have to cut off the weak
localization corrections [ $\propto \alpha_{\rm wl}$ in expression
(\ref{eq:RtH})] by $T_H= e H D /\hbar c$.

\subsection{Definition of transresistances}
\label{se:defs}
Now we are at the position to define transresistances. As was
demonstrated, $U_x$ and $U_y$ present a generalization of transvoltage
and Hall transvoltage that are integrated on the boundary. In order to
define the transresistances properly we divide $U_i$ by the current
injected from the leads to layer 1 and by the appropriate
circumference. This gives the following definition:
\begin{equation}
\label{def:RtxRty}
R_{ti} = \frac{U_i}{ I \oint _{\partial S_2} \left|n_i\right| dl/2}
       = \frac{-\int \!\!\! \int_{S_2} \nabla_i
V^{(2)}(x,y) dx dy}{J W_1 \oint_{\partial S_2} | n_i| dl/2}
\end{equation}
A rectangular shape of both layers (of size $L_1 \times W_1$ and $L_2
\times W_2$) reduces these definitions to:
\begin{eqnarray}
\label{def:Rt}
R_{tx} \equiv R_t=\frac{U_x}{J W_1 W_2} \;\;\;\;\;\;\;\;\;\;\;\;\;\nonumber\\
= \frac{1}{W_2 I} \int_{-W_2/2}^{W_2/2} dy \left[ V^{(2)}(-L_2/2,y)-
V^{(2)}(L_2/2,y) \right]
\end{eqnarray}
and 
\begin{eqnarray}
\label{def:RtH}
R_{ty} \equiv R_t^H=\frac{U_y}{J W_1 L_2} \;\;\;\;\;\;\;\;\;\;\;\;\; \nonumber \\
=\frac{1}{L_2 I} \int_{-L_2/2}^{L_2/2} dx\left[ V^{(2)}(x,-W_2/2)-
V^{(2)}(x,W_2/2) \right].
\end{eqnarray}
(The integration over the edges can be done experimentally by
connecting metallic measuring probes through high barriers.)
Using the relation ${\vec U}={\vec P}+{\vec F}$ [see Eq.~(\ref{eq:U})]
and expressions (\ref{eq:Pc}) for ${\vec P}$ and (\ref{eq:F}) for
${\vec F}$, we find the expressions for the transresistance
(\ref{eq:Rt}) and the Hall transresistance (\ref{eq:RtH}) cited in the
Introduction.

\section{Microscopic theory of $\bbox{\sigma}$, $\bbox{\sigma_H}$, $\bbox{\sigma^D}$,
  $\bbox{\sigma^D_H}$, \lowercase{$\bbox{g^l}$}, $\bbox{{\sf
      Q}^{\lowercase{l}}}$ and $\bbox{{\sf Q}^{\lowercase{l}}_H}$}

\label{se:micro}

In the section we review the dependence of the microscopical,
geometry--independent parameters $\sigma$, $\sigma_H$, $\sigma^D$,
$\sigma^D_H$ and $g^l$ on the temperature, the interlayer tunneling
amplitude, the level of disorder in each layer, and the strength of
the intralayer and interlayer interaction. We derive the behavior of
${\cal Q}_{ij}^{\lowercase{l}} ({\vec r}, {\vec r'})$ introduced in
Eq.~(\ref{def:tilsigma})and derive ${\sf Q}^{\lowercase{l}}_{ij}$
defined in Eq.~(\ref{eq:Qij0}).

\subsection{Review : \bbox{$g^l$, $\sigma$, $\sigma_H$} and \bbox{$\sigma^D$}} 

\label{sse:rev}

The effect of disorder and Coulomb interaction in a single 2D layer
was studied intensively at the beginning of the 1980s
\cite{DS:Altshuler82}.  The interplay between weak disorder,
interference and Coulomb interaction leads to various logarithmic
corrections to the sheet conductance $\sigma$ and to the Hall
conductance $\sigma^H$.  It was found that \cite{DS:Altshuler82}:
\begin{equation}
\label{eq:sig}
\sigma = \left(1/R_\square \right) \left[ 1 - t_\square (\alpha_{\rm int}+
\alpha_{\rm wl}) \log( 1/ \tau T) \right]
\end{equation}
where $t_\square = R_\square e^2/2 \pi^2 \hbar$, and the $O(1)$
dimensionless coefficients, $\alpha_{\rm int}$ and $\alpha_{\rm wl}$,
describe the corrections due to interaction and weak localization
respectively.
 
The parameters $g^l$ describe tunneling between the two layers. The
finite conductance inside each layer reduces the speed of the charge
spreading after the electron tunnels, and leads to a suppression of
the tunneling rate \cite{DS:Altshuler79,DS:Levitov96}.  To first order
in the intralayer $e$--$e$ interaction
\begin{equation}
\label{eq:tl}
g^l \Rightarrow g^l \left[ 1 - t_\square \beta_{\rm zba} \log( 1/ \tau
 T) \right],
\end{equation}
where $\beta_{\rm zba} = \log(\kappa d)$, $\kappa= 2 \pi e^2 \nu $ is
the inverse Thomas - Fermi screening radius.
There are no interaction corrections to $\sigma^H$, however weak
localization corrections exist and they are twice as larger as the
weak localization corrections to the conductance,
\begin{equation}
\label{eq:sigH}
\sigma^H \Rightarrow \sigma^H 
\left[ 1 - 2 t_\square \alpha_{ \rm wl} \log(1 / \tau T)\right].
\end{equation}
They tend to be suppressed by a weak magnetic field, however in the
limit when the magnetic length $\sqrt{c / 2 e H }$ is larger than the
dephasing length ( that is proportional to the temperature in 2D) they should
be included.

The expressions for $\sigma^D$ and $\sigma_H^D=2 \sigma_H \sigma_D/
\sigma$ are given in Eqs.~(\ref{eq:sigmaD}) and (\ref{eq:rhoD}).  They
were derived in that form in Eqs. (35) and (26) of
Ref.~\cite{DR:Kamenev95}.

\subsection{The matrix ${\cal Q}^{l}_{ij} ({\vec r}, {\vec r'})$}

\label{sse:calQ}

By definition the matrix ${\cal Q}^{l}_{ij} ({\vec r}, {\vec r'})$ is
given (using a linear response formalism) by the correlation function
of the currents $J^{(1)}_i({\vec r})$ with $J^{(2)}_j({\vec r'})$.
The Hamiltonian of the system is given by \cite{DR:Oreg98}
\begin{equation}
H=H_{1}+H_{2}+H_{\mbox{\footnotesize int}}+H_{T}\,,  \label{H}
\end{equation}
where $H_{1(2)}$ is the Hamiltonian of the isolated layer 1 (2),
including elastic disorder, and $H_{\mbox{\footnotesize int}}$
includes interlayer as well as intralayer Coulomb interactions. The
first three terms on the r.h.s. of Eq.~(\ref{H}) are traditionally
involved in the description of the drag effect
\cite{DR:Zheng93,DR:Kamenev95,DR:Flensberg95}. We add the term
describing pointlike tunneling processes through pinholes
\cite{DR:foot5}
\begin{equation}
H_{T} = \sum\limits_{l=1}^{N} V^l \sum\limits_{{\vec k},{\vec p}} e^{i
\vec R^l \cdot ({\vec k}-{\vec p})} a_{\vec k}^{\dagger}b_{\vec p} +
\mbox{h.c.}  \, , \label{HT}
\end{equation}
where $a(a^\dagger)$ and $b(b^\dagger)$ are the annihilation
(creation) operators of electrons in the first and the second layers
respectively, and ${\vec R}^l$, $l=1\ldots N$, are the positions of
$N$ local pinholes.  The coupling energy $V^l$ is related to the
tunneling amplitude $t^l$ by $t^l = \pi \nu \sqrt {2 L_1 W_1 L_2 W_2}
V^l$.

The actual calculations of current--current correlation functions are
very similar to the calculations of interaction corrections to the
conductivity and the Hall coefficient in a single layer. In the
description of the calculation here step by step we will give
references to the equivalent steps in case of single layer calculation
that are described in some details in Ref. \cite{DS:Altshuler80}.

\subsubsection{Diagonal elements: ${\cal Q}^{l}_{xx} ({\vec r}, {\vec r'})$}

\label{se:Qii}

\begin{figure}[h]
\vglue 0cm \hspace{0.1\hsize} \epsfxsize=1 \hsize
\epsffile{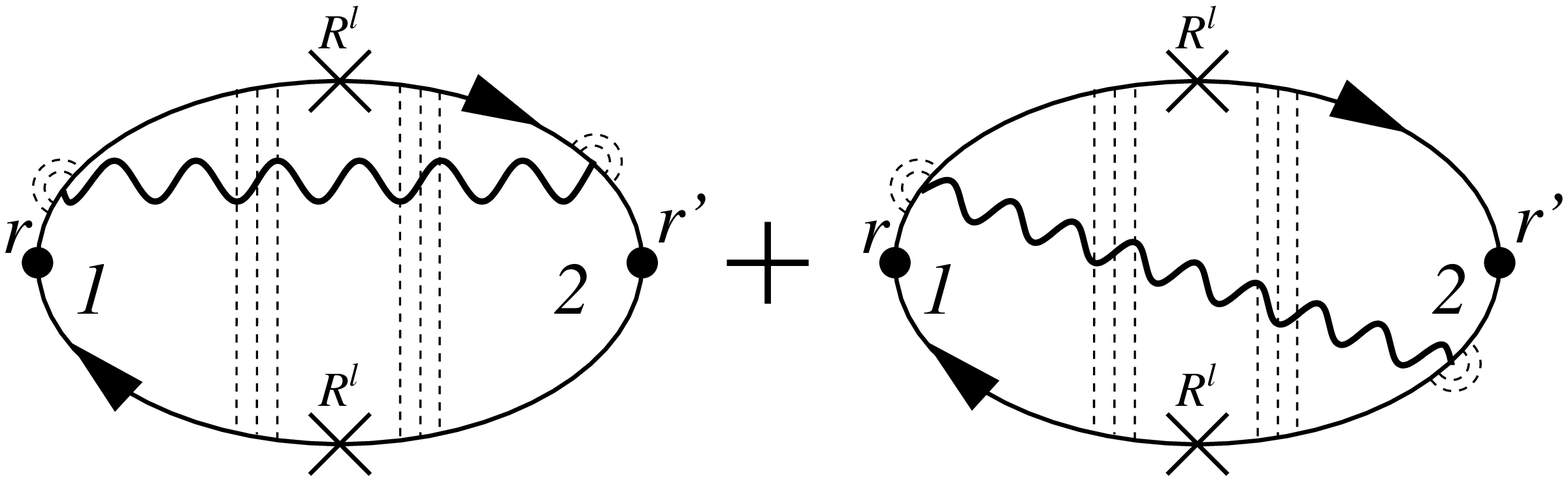}
 \refstepcounter{figure} \label{fg:Qrr} \\
 {\small FIG.\ \ref{fg:Qrr} Two diagram describing the
   transconductivity matrix ${\cal Q}^{l}_{ii}$ that are second order
   in interlayer tunneling (denoted by $\times$)} (two additional
 diagrams with opposite arrows should be included). Solid lines with
 arrows are electrons propagators. Full circles denote current vector
 vertexes at point $\vec r$ in layer 1 where the external electric
 field is applied and at point $\vec r'$ in layer 2 where the current
 is measured.  The crosses denote tunneling through a pinhole located
 at ${\vec R}^l$.  Since impurity (dashed) lines are local on scale of
 the mean free path, the interlayer interaction (wavy) line is of
 range of the order of the distance between the layers $d$, and the
 relevant frequencies in the diffuson (group of dashed lines) are of
 the order of the temperature $T$, ${\cal Q}_{xx}^l( {\vec r}, {\vec
   r'})$ decays strongly when the distance between ${\vec R}_l$ and
 ${\vec r}$ or ${\vec r'}$ is larger than the thermal length $L_T=
 \sqrt{ D/ T}$.
\end{figure}

The largest contributions to ${\cal Q}^{l}_{xx} ({\vec r}, {\vec r'})$
(to second order in tunneling amplitude, and first order in interlayer
interaction) is coming from the diagrams depicted in
Fig.~\ref{fg:Qrr}, (two additional diagrams where the direction of the
arrows is inverted should be added). These diagrams are equivalent to
the``three diffuson'' diagrams contributing to the first order
intralayer interaction corrections to the conductivity of a single
layer.  [Diagrams 5 (d) and (e) of Ref.~\cite{DS:Altshuler80}.] In the
present case, due to the tunneling between the layers, the corrections
(to second order in tunneling amplitude) are due to interlayer
interaction and there are two diffusons in each layer, i.e., four
diffusons in total, and we expect more singular temperature
dependence.

These diagrams are calculated using the Matsubara frequency formalism
\cite{RFS:Mahan90}.  The analytical structure of the frequencies in
them is identical to the structure in case of single layer interaction
corrections to the conductivity \cite{DS:Altshuler80}. In case of a
single layer there are, in addition to diagrams equivalent to the ones
in Fig~\ref{fg:Qrr}, ``two diffuson'' diagrams that eventually cancel
each other out.  [Diagrams 5 (a) (b) and (c) in
Ref.~\cite{DS:Altshuler80}.] This cancellation involves a single
impurity line [Diagram 5 (c) in Ref.~\cite{DS:Altshuler80}] that is
not present here, because by assumption there are no common impurities
(besides the tunneling impurity) to the two layers. However, in the
present case, due to the integration over the different fast momenta
in the current vertexes and the singular behavior of the four diffuson
diagrams, the contribution of two diffuson diagrams is less singular
and can be neglected.

After integration over the fast momenta, the formal expression for the
longitudinal term ${\cal Q}^l_{xx}(\vec r - \vec R^l,r'- R^l)$
associated with the diagrams in Fig.~\ref{fg:Qrr} reduces to
\cite{DR:Oreg98}:

\begin{eqnarray}
{\cal Q}^l_{xx}(\vec r - \vec R^l, \vec r' - \vec R^l)=
\;\;\;\;\;\;\;\;\;\;\;\; \nonumber \\
i \frac{\sigma}{4\pi} \int\limits_{-\infty }^{\infty}\!\!\!\!d\omega
\frac{\partial }{\partial \omega } \left[ \omega \coth \frac{\omega
}{2T} \right]
F_{xx}(\vec r- \vec R^l,r'-\vec R^l;\omega )\,.
\label{AA1rr}
\end{eqnarray}
The function $F_{xx}(\vec r - \vec R^l ,\vec r' - \vec
R^l;\omega )$ to second order in the tunneling amplitude is given by:
\begin{eqnarray}
\label{eq:Fa}
F^{\rm (a)}_{xx}(\vec r - \vec R^l ,\vec r' - \vec R^l;\omega )=
8 \frac{R_\square}{R^l_\perp} \frac{D}{(L_1 W_1 L_2 W_2)} 
\times \nonumber \\
\sum_{Qkq}D
(Q_x + q_x/2)(Q_x + k_x/2) U_{12}(Q,\omega) \times \nonumber \\
{(DQ^{2}-i\omega)^{-2}}[(D(\vec Q + \vec k)^{2}-i\omega)(D(\vec Q +
\vec q)^{2}-i\omega)]^{-1}  \times \\
e^{i\vec q \cdot (\vec r - \vec R^l)} e^{-i
\vec k \cdot (\vec r' - \vec R^l)} \, , \nonumber
\end{eqnarray}
where $R_\perp^l=2\pi \hbar/ e^2 |t^l|^2$ is the resistance due to
tunneling through the pinhole located at ${\vec R}^l$, and
$U_{12}(Q,\omega)$ is the interlayer screened Coulomb interaction,
which is given in the diffusive case, by \cite{DR:Kamenev95}:
\begin{equation} 
\label{eq:intDif}
U_{12}(\vec q, \omega )= 
\frac{1}{S_{\rm int}}\frac{\pi e^2 q}{\kappa^2 \, \sinh qd} 
\left(\frac{D q^2 - i \omega}{D q^2}\right)^2,  
\end{equation}
where the divergences at small $q$ are cut off at 
$D q^2 \approx \omega/(\kappa d)$.

Performing the integration over $Q$ we find 
\begin{equation}
\label{eq:intQ}
\int d^2 s d^2 s' {\cal Q}^l_{xx}(\vec s, \vec s') = 
-\frac{e^2}{\hbar} \frac{1}{24 \pi}\frac{\ln (\kappa d) }{\kappa d}
\frac{R_\square}{R_\perp^l} L_T^2.
\end{equation}
On the other hand since the frequencies are of the order of the
temperature and the momenta $Q q k$ are of the order of $\sqrt {D/
  \omega}$ the function $F^{\rm a}$ and therefore ${\cal Q}^l$ decays
at range of order of $L_T$.  Physically, the relevant time for the
quantum phenomena under discussion has to be smaller than $\hbar /T$.
At time $\propto 1/T$ the electron diffuses for a length $L_T$, hence
the quantum correction has a range $L_T$.  This allows us to write
${\cal Q}^l$ as
\begin{equation}
\label{eq:Ql}
 {\cal Q}^l_{ij}({\vec r}, {\vec r'}) \cong {\sf Q}^l_{ij} \delta({\vec
r} -{\vec r'}) S_T^l(|{\vec r'}|)
\end{equation}
where $S_T^l(\vec r)$ is a function of range $L_T$ around ${\vec
R}^l$.  We normalized it in such a way that $\int d^2 r S_T^l(\vec r)
= L_T^2$.  Integrating the right hand side of Eq.~(\ref{eq:Ql}) with
respect to $\vec r$ and $\vec r'$ on all space we find that:
\begin{equation}
\label{eq:sfQ}
\int {\sf Q}_{ij}^l \delta (\vec s - \vec s') S^l_T(|\vec s|) d^2 s d^2 s' = L_T^2 {\sf Q}_{ij}^l
\end{equation}
Using Eqs.~(\ref{eq:intQ}-\ref{eq:sfQ}, ) we find:
\begin{eqnarray}
\label{eq:sfQa}
{\sf Q}=\sum_l{\sf Q}_{xx}^l= \sum_l \frac{1}{L_T^2} \int d^2 s d^2
 s' {\cal Q}^l_{xx}(\vec s, \vec s') = \nonumber \\ -\frac{e^2}{\hbar}
 \frac{1}{24 \pi}\frac{\ln (\kappa d) }{\kappa d}
 \frac{R_\square}{R_\perp},
\end{eqnarray}
where $1/R_\perp = \sum_l 1/R_\perp^l$. 

The formulas assume that $L_T$ is small compared to the system
dimensions $L$ and $W$. When the temperature $T$ becomes so low that
$L_T$ is greater then both $L$ and $W$ the transresistance should be
flatten out and become independent on the temperature. For a system
where $L \gg W$, there should also be an intermediate temperature
regime $L > L_T > W$, where the system becomes
quasi--one--dimensional, the integration over the momenta in
Eq.~(\ref{eq:Fa}) are one dimensional and the divergences are more
singular in the temperature.  In that case integration over the
momenta in Eq.~(\ref{eq:Fa}) [assuming that $L_{\min} > d$ and
therefore the screening properties have a $2D$ character] gives:
\begin{equation}
\label{eq:Q1D}
{\sf Q} = - c \frac{e^2}{2 \pi \hbar} \frac{1}{\sqrt {\kappa d}}
\frac{R_\square}{R_\perp} \frac{L_T}{L_{\min}},
\end{equation}
where $c= 1/\left(2 \pi^{7/2} \right) \zeta\left(5/2 \right) \sim
1/13$.

In addition to the exchange diagrams depicted in Fig.~\ref{fg:Qrr}
diagrams of Hartree type should be included.  However, unlike the
situation in a single layer system \cite{DS:AltLee80}, the
contribution of Hartree type diagrams is not singular as the
contribution of the exchange diagrams of Fig.~\ref{fg:Qrr}. Formally
it happens because the Coulomb line in a Hartree diagram does not
transfer momenta between the layers.

Let us examine now what happens for higher order terms, i.e., when we
have more then a single event of interlayer tunneling.  For two
interlayer tunneling events (fourth order in the tunneling amplitude)
at point ${\vec R}^l$ and ${\vec R}^{l'}$, we have to consider several
diagrams.  A typical contribution of them is the term:

\begin{eqnarray}
\label{eq:Fb}
F^{\rm (b)}_{xx}({\vec r} - {\vec R}^l ,{\vec r}'- {\vec R}^l;{\vec r} - {\vec
R}^{l'} ,{\vec r}' - {\vec R}^{l'}\omega ) =\nonumber \\
8 \left[\frac{R_\square}{R_\perp} \frac{D}{\sqrt{L_2 W_2 L_1 W_1}}\right]^2
\frac{1}{L_2 W_2} \times \nonumber \\
\sum_{PQkq}D (Q_x + q_x/2)(P_x + k_x/2) U_{11}(Q,\omega)
\times  \nonumber \\
{(DQ^{2}-i\omega)^{-2}} [(D({\vec Q} + {\vec q})^{2}-i\omega)
(DP^{2}-i\omega)]^{-1} \times \\
(D(\vec P + \vec k)^{2}-i\omega)^{-1} \times \nonumber \\
e^{i {\vec q} \cdot ({\vec r} - {\vec R}^l)} e^{-i {\vec k} \cdot (\vec r' -
{\vec R}^l)}e^{ -i ({\vec Q} - {\vec P}) \cdot {\vec R}^l} e^{ i (\vec Q -
{\vec P}) \cdot {\vec R}^{l'}} \, . \nonumber
\end{eqnarray}
If we assume that the tunneling is at one point than integration over
the direction of the momentum $\vec P$ causes the integral to vanish.
When we average on the position of the tunneling points at an area $a
\times b$ we find for $a \ll L_T$ that the result is less singular
than $1/T$ and for $a> L_T$ we recover the results 3(b) and (15) of
Ref.~\cite{DR:Oreg98} in which ${\sf Q} \propto 1/T^2 \log T$.  Other
diagrams give similar results.

Practically, if we are interested in situations when the quantum
contribution is dominant, we are interested in the case $a \ll L_T$,
and we can ignore the second order contribution.
\begin{figure}[h]
\vglue 0cm \hspace{0.1\hsize} \epsfxsize=1 \hsize
\epsffile{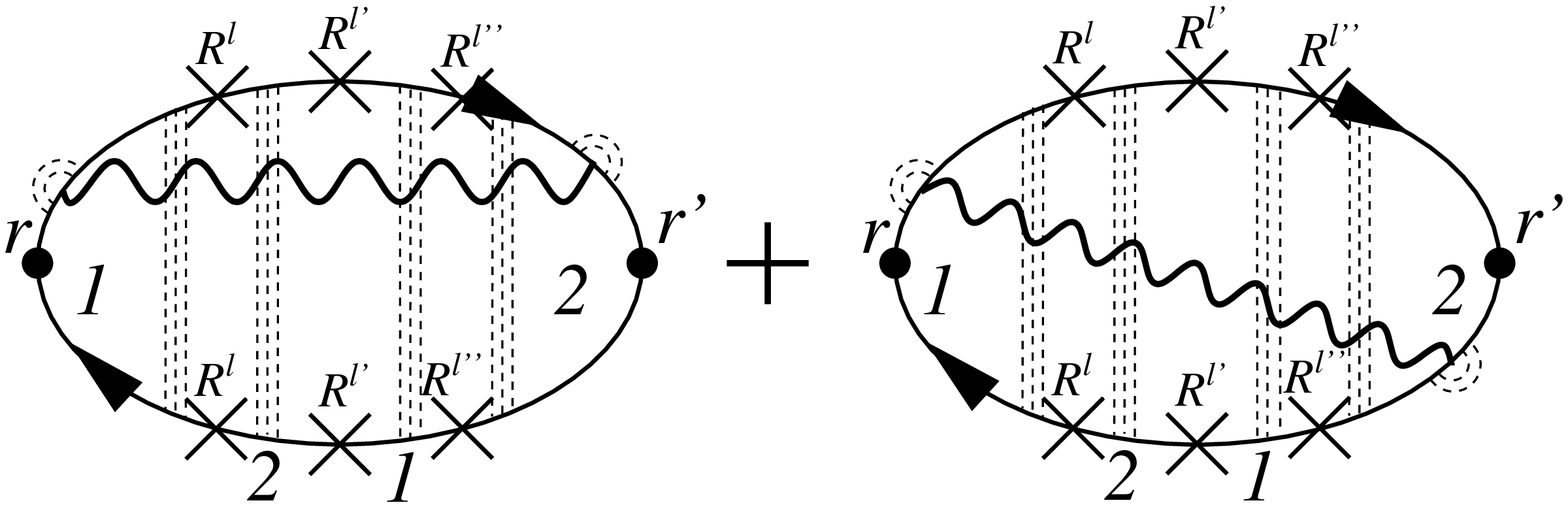}
 \refstepcounter{figure} \label{fg:Qrr3} \\
 {\small FIG.\ \ref{fg:Qrr3} The most divergent diagrams contributing
   to the third order tunneling (6th order in tunneling amplitude)
   transconductance.  Two additional diagrams with opposite arrows
   should be included.}
\end{figure}

The third order contributions, involving three tunneling events (6th
order in tunneling amplitudes), may also be computed. The most
divergent contributions, depicted in Fig.~\ref{fg:Qrr3}, are found by
inserting four additional crosses in the diagrams of Fig.~\ref{fg:Qrr}
in a way that gives the maximal number of diffusons.  For the case $a
\ll L_T$ we find that their contribution to ${\sf Q}$ is
\begin{equation}
\label{eq:3order}
{\sf Q}^{\rm (c)}= {\sf Q}^{\rm (a)} \left(\frac{R_\square} {\pi R_\perp}
\log {\frac{1}{T \tau}} \right)^2,
\end{equation}
where ${\sf Q}^{\rm (a)}$ is the first order contribution to {\sf Q}
given by (\ref{eq:sfQa}). Thus, the first order results are valid as
long as $[R_\square/ (\pi R_\perp)] \log (1/T \tau) < 1$.  In the quasi
1D case the condition is $(R_\square/ R_\perp)(L_T/ L_{\min}) \ll
1$.

By examining carefully the analytical structure of the diagrams in
Fig.~\ref{fg:Qrr}, we can try to understand the physics leading to the
singular quantum contribution (\ref{eq:sfQa}) to the transconductance.

The ground state of the interacting system may be built up from
virtual particle--hole ($p$--$h$) excitations, relative to the Fermi
sea of the non--interacting disordered system. The $e$--$e$
interaction, at some instant of time, can create two $p$--$h$ pairs,
which propagate forward in time; it can annihilate two $p$--$h$ pairs
which originated at an earlier time; it can annihilate one $p$--$h$
pair and create another; it can lead to scattering among existing
holes and/or particles; or it can create or annihilate a single pair
while scattering an existing particle or hole.  The most commonly
considered correction to the groundstate energy of the Fermi liquid,
beyond the Hartree--Fock, is the RPA correction, which is represented
diagrammatically by closed chains of two or more $p$--$h$ bubbles.  In
these diagrams, each particle is annihilated by the same hole that was
created with it originally, and there are no scattering processes for
existing electrons or holes.

If one ignores the momentum vertices, the diagram in Fig.~\ref{fg:Qrr}
is a contribution to the RPA ground state energy in which one of the
$p$--$h$ pairs initially created by the Coulomb interaction at a time
$t_0$, tunnels at from one layer to the other, before being
annihilated at a later time $t_f$.  The second $p$--$h$ pair does not
tunnel between the two layers, but forms part of a polarization cloud,
represented by the screened interaction propagator, which propagates
in both layers because of the interlayer Coulomb interaction.

Now let us insert momentum vertices in the diagram, as shown in Fig 2,
so that we can compute the time-dependent correlation function between
the momenta in the two different layers.  From this correlation
function, using the Kubo formula, one may compute the
transconductance.  If the $p$--$h$ pair is created at time $t_0$ with
total momentum $q$, the polarization propagator will carry wavevector
$-q$.  Although the system contains impurities, so that the electron
momentum is not conserved, the diagrams which include averaging over
impurity positions preserve the total momentum of the $p$--$h$ pair,
and of the screened electron propagator.  The tunneling processes
randomize the momenta of the particle and hole, when they cross from
one layer to the other.  Consequently, in the absence of Coulomb
interactions, there would be no correlation in the total momenta of
the two layers.  Now however there is an induced correlation: in order
for the $p$--$h$ pair to be annihilated at time $t_f$ by the
polarization propagator created at time $t_0$, it must have the same
total momentum $q$ as the original pair.

The overall sign of the correlation depends on the sign of the
interlayer $e$--$e$ interaction.  Indeed, if one changes the sign of
the interlayer $e$--$e$ layer, while keeping the fixed the interaction
between electrons in the same layer, the contribution of the diagrams
in Fig.\ref{fg:Qrr} will simply change sign. (This follows from the
fact that the bare interlayer interaction appears an odd number of
times in the screened interaction propagator for this case.) For the
actual situation, where the interlayer interaction is repulsive, we
find a negative contribution to the transconductance.

When the layers are in the diffusive regime, the particle and hole of
an excitation stay closer to each other in position space, compared to
the ballistic case, and stay closer to their initial position.  This
increases the probability that they tunnel through the same pinhole,
and that they eventually annihilate the polarization cloud that was
originally created together with them.  In the diagrams, this effect
is presented by the dressing of the vertexes by the diffuson
propagators.

Our analysis shows that the correlation of the momentum fluctuations,
at $T=0$, extends for very long times, giving rise to a singular
contribution to the transconductance.  At finite temperatures, when
the discontinuity at the Fermi energy is rounded, the contribution of
the virtual processes will be cut off at times of order $1/T$, and the
singular contribution is reduced.
\subsubsection{The Hall coefficient ${\cal Q}^{l}_{xy} ({\vec r}, {\vec r'})$}
\label{se:Qxy}
\begin{figure}[h]
  \vglue 0cm \hspace{0.05\hsize} \epsfxsize=1\hsize
  \epsffile{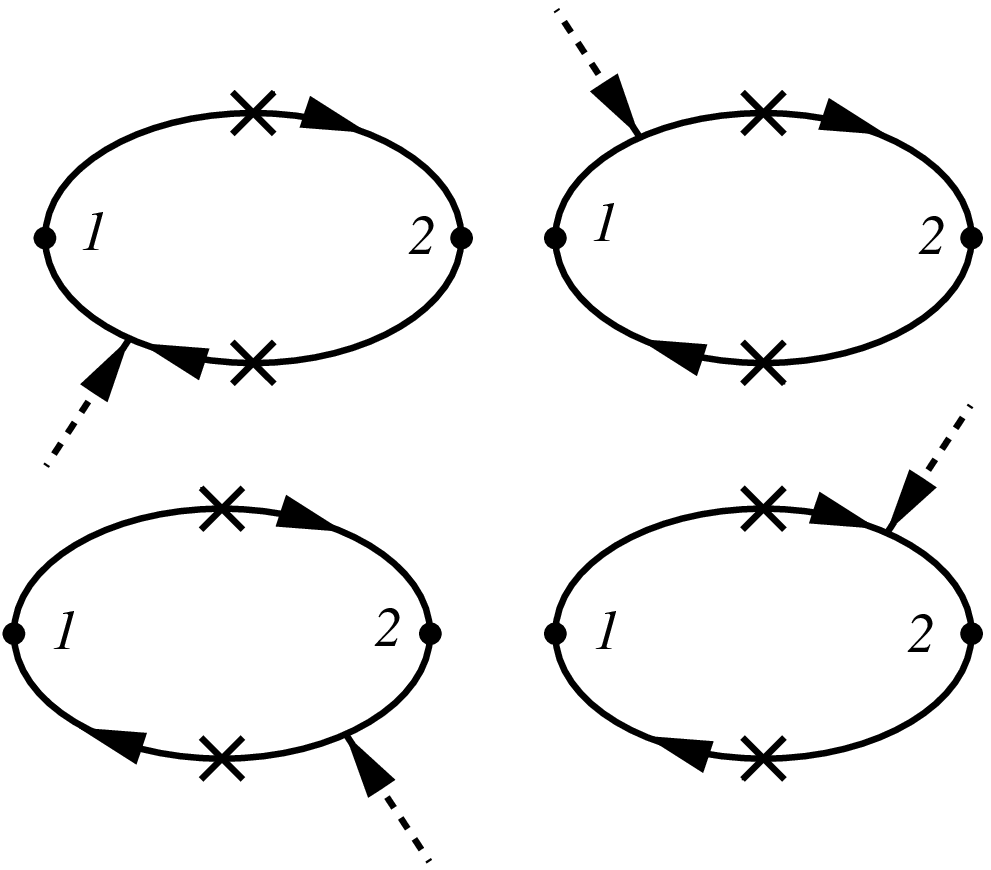}
  \refstepcounter{figure} \label{fg:Hdni} \\
 {\small FIG.\ \ref{fg:Hdni} Diagrams describing the Hall
   transconductivity that are second order in interlayer tunneling
   (denoted by $\times$) and does not include $e$--$e$ interaction.
   Dashed vector signifies an external magnetic field, solid lines are
   electron propagators and the full dots are current vertexes. An
   average over the current direction at the vertexes shows that this
   contribution vanishes.}
\end{figure}

In the absence of a magnetic field and without any $e$--$e$
interactions the electrons' wave front propagating in layer 2 (after
tunneling from layer 1) is spherically symmetric~\cite{DR:foot5}.
This argument is not changed in the presence of an external magnetic
field, (though the current distribution has now a component
perpendicular to the radial direction), and leads us to the conclusion
that {\it without} $e$--$e$ interaction the tunneling itself between
the layers does not lead to a finite Hall transconductivity
coefficient.  Generalizing the calculation of the Hall coefficient in
a single layer [see Figs. 2 (a) and (b) in Ref.~\cite{DS:Altshuler80}]
to the case of two layers we find that the diagrams describing the
Hall transconductivity coefficient, ${\cal Q}^l_{xy}={\cal Q}^l_{H}$,
without $e$--$e$ interactions, are the ones depicted in
Fig.~\ref{fg:Hdni}.  An average over the current direction in the
vertexes (full dot in Fig.~\ref{fg:Hdni}) proves that this
contribution vanishes. The insertion of Cooperons [compare to Fig 3.
in Ref.~\cite{DS:Altshuler80}] will not change that result.

\begin{figure}[h]
\vglue 0cm \hspace{0.05\hsize} \epsfxsize=1\hsize
\epsffile{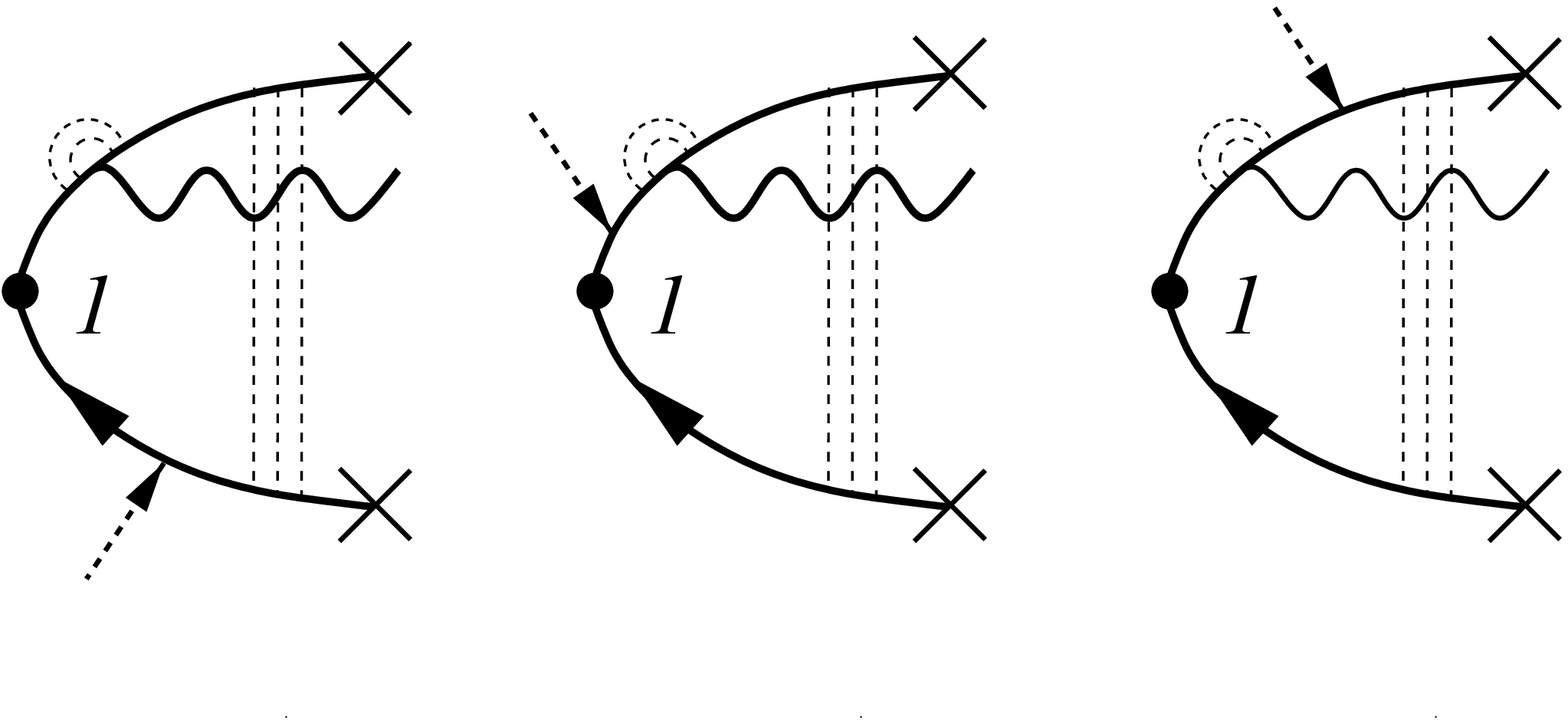} 
\refstepcounter{figure} \label{fg:Hdi} \\
{\small FIG.\ \ref{fg:Hdi} The different ways to insert a magnetic
  vertex (dashed line with an arrow) to the left part of the
  conductivity corrections depicted in \protect{\ref{fg:Qrr}}. These
  diagrams are similar to those of Fig.~6 in
  \protect{\cite{DS:Altshuler80}}. The sum of these diagrams is zero.
  The contribution of ``two diffuson'' diagrams, (without the diffuson
  connecting the two sides of the current vertex) vanishes as well.}
\end{figure}

We shall now show that that inclusion of interlayer Coulomb
interactions does not affect the results that the tunneling give no
contribution to the Hall transconductivity.  As in the case without
interaction, we insert a magnetic vertex in all possible ways to the
diagrams contributing to the transconductivity. Concentrating first in
the part related to layer 1 of the diagrams depicted in
Fig~\ref{fg:Qrr} we find three possible ways to enter the magnetic
vertex as shown in Fig.~\ref{fg:Hdi}.

These diagrams are similar to those of interaction corrections to the
Hall coefficient in a single layer [figure 6 of
Ref.~\cite{DS:Altshuler80}].  It was shown in
Ref.~\cite{DS:Altshuler80} that they cancel each other out.

We showed that there is no noninteracting Hall transconductivity and
that the most singular terms contributing to the transconductivity do
not give rise to any Hall transconductivity as well.  Since we look
for a term different from zero we have to check also that the ``two
diffuson'' diagrams does not contribute to the Hall transconductivity.
But, the insertion of a magnetic field vertex to the two diffuson
diagrams gives three diagrams similar to the one depicted in figure
\ref{fg:Hdi}. For the ``two diffuson'' diagrams the diffuson
connecting the two parts of the current vertex is missing and the
Green functions in the two sides of the current vertex has opposite
imaginary part. It can be shown, however, that in this case as in the
case of the diagrams in Fig.~\ref{fg:Hdi} the sum of the three
vanishes. The calculation is almost identical to the calculation in
Ref.~\cite{DS:Altshuler80} and we will not repeat it here.

This leads us to the conclusion that to first order in interaction
\begin{equation}
\label{eq:sQH}
{\cal Q}_{xy}^l={\cal Q}^l_H={\sf Q}^l_H=0
\end{equation}

This implies, in turn, that the quantum process gives no contribution
to the Hall trans{\it conductance}. When the conductivity matrix is
inverted, however, this leads to a {\it nonvanishing} contribution to
the Hall trans{\it resistance}, following the analysis of
Sec.~\ref{se:F}.

\section{The potential in layer $\bbox{2}$: $\bbox{V^{(2)}\lowercase{(x,y)}}$}

\label{se:V2}

In the previous sections we have discussed the relation between
integrated potentials measured on the edges of layer 2 and the applied
current in layer 1. We have used Gauss's theorem and did not need to
calculate the potential in layer 2, $V^{(2)}(x,y)$ in details.
Nevertheless, it is instructive to understand, in few simple examples,
the detailed behavior of $V^{(2)}(x,y)$. In principle, with advanced
technology, $V^{(2)}(x,y)$ can be measured directly experimentally
\cite{QHE:Ashoori98}. We will discuss the two situations mentioned in
the introduction: the ``parallel strip'' and ``cross'' geometries with
pinhole distribution that is concentrated in the middle of the sample,
and identical layers with pinholes that are distributed uniformly.  In
all cases we assume that the 2D gases have rectangular shapes of sizes
$L_1 \times W_1$ and $L_2 \times W_2$.

\subsection{Tunneling through pinholes in the middle of the sample }

\label{se:middle}

Now we assume that there are $N$ pinholes distributed inside a
rectangular of size $a \times b$ at the origin and $a, b \ll
L_1,L_2,W_1,W_2$. For simplicity, as before, we assume that the
frictional forces and the tunneling between the layers are weak, i.e.,
$\rho_D \ll R_\square$ and $R_\square \ll R_\perp $. In that case we
can solve the equations (\ref{eq:con}), (\ref{eq:ohmslawf}) with the
boundary condition (\ref{eq:bc12}) perturbatively in
$\rho_D/R_\square$ and $R_\square/R_\perp$.

If $W_2 \le W_1$ and $L_2 \le L_1$ then the classical drag is present
all over layer $2$ and its contribution to the potential in layer $2$
can be readily found by changing Eq.~(\ref{eq:J0}) to:
\begin{equation}
\label{eq:J0c}
{J_i}_{(0)}^\alpha ({\vec r}) = -\left[\sigma_{ij} \delta_{\alpha \beta} +
\sigma_{ij}^D X^{\alpha \beta} \right] \nabla_j V_{(0)}^\beta(\vec r) ,\;\;\;
\end{equation}
where the diagonal elements of the matrix $X^{\alpha \beta}$ are zero
and the off diagonal are $1$.  Substitution in Eq.~(\ref{eq:con})
(with vanishing right hand side at the discussed case) and using
(\ref{eq:contenb}) we find the Laplace equation $\Delta
V_{(0)}^\alpha=0$ with the boundary condition (\ref{eq:bc12}).  The
solution is straight forward and in the limit of $\sigma_D \ll \sigma$
given by:
\begin{equation}
\label{sol:V0D}
V_{(0)}^{(1)}= - J\frac{\sigma}{\sigma^2+\sigma_H^2} x - J
\frac{\sigma_H}{\sigma^2+\sigma_H^2} y,\;\;\; V_{(0)}^{(2)}= J \rho_D x
\end{equation}
Notice that the Hall transvoltage vanishes. This happens due to the
relation (\ref{eq:sigmaD}),\cite{DR:Kamenev95}.

In order to find (perturbatively) the voltage in layer 2 in the
presence of tunneling we substitute Eqs.~(\ref{def:V1}) and
(\ref{eq:Ql}) in Eq.~(\ref{eq:ohmslawf}), keeping terms up to first
order in $R_\square / R_\perp$ and $\rho_D/R_\perp$ we find:
\begin{eqnarray}
\label{eq:J1}
{J}_i^\alpha ({\vec r}) = {J_{(0)}}_i^\alpha(\vec r) + & \sigma_{ij}
\delta_{\alpha \beta} & \bbox{\nabla}_j V_{(1)}^\beta({\vec r})
\nonumber \\ +& \sum_l{\sf Q}^l_{ij} S_T^l(|{\vec r}|)X^{\alpha \beta } &
\bbox{\nabla}_j V_{(0)}^\beta ({\vec r}).
\end{eqnarray}
By assumption the corrections to the currents in layer 1 are small and
we neglect them. A substitution in the continuity equation
(\ref{eq:con}) and a use of (\ref{eq:contenb}) yields the following
equation for $V_{(1)}^{(2)}$, representing the correction to the
voltage in layer 2 due to tunneling
\begin{eqnarray}
\label{eq:V1}
-(\sigma+\sigma^D) \Delta V_{(0)}^{(2)} -\bbox{\nabla}_i \sigma_{ij}
 \bbox{\nabla}_j V_{(1)}^{(2)} = \nonumber\\
 -g^t V^{(1)}_{(0)} -
 \bbox{\nabla}_i \sum_l{\sf Q}_{ij}^l S_T^l(|{\vec r}|) \bbox{\nabla}_j V^{(1)}_{(0)}
\end{eqnarray}
The first term in the left hand side of that equation vanishes, and we
are left with an equation for $V^{(2)}_{(1)}$. The boundary condition
for $V^{(2)}_{(1)}$ should be such that the residual current due to
the potential $V^{(2)}_{(1)}$ vanishes on all boundaries, i.e,
\begin{eqnarray}
\label{eq:bcv12}
{J_{(1)}^{(2)}}_x &=& \sigma \nabla_x V_{(1)}^{(2)}(\pm L_2/2,y) \nonumber \\
& &+\sigma_H \nabla_y V_{(1)}^{(2)} (\pm L_2/2,y) = 0 \nonumber \\
 {J_{(1)}^{(2)}}_y&=& -\sigma_H \nabla_x V_{(1)}^{(2)} (x,\pm W_2/2)  \nonumber \\
& &+ \sigma \nabla_y V_{(1)}^{(2)} (x,\pm
W_2/2)= 0
\end{eqnarray}
Using (\ref{eq:contenb}) for the intralayer conductivities we end up
with a Poisson equation:
\begin{equation}
\label{eq:So}
\sigma \Delta V^{(2)}_1 = {S},
\end{equation}
where the source term, ${S}$, is $(-)$ the right hand side of
Eq.~(\ref{eq:V1}).

The Green function of this equation is given by the solution of
Eq.~(\ref{eq:So}) and boundary condition (\ref{eq:bcv12}) with a
source term equal to $ \delta(\vec r-\vec r')$. In the absence of
magnetic field the boundary conditions can be easily satisfied, by
introducing a series of image charges. Since the boundary conditions
are of Neumann type, all the charges have the same sign.  In the
presence of magnetic a field we should add to the image charges a
source of circulation that will cancel the twist of the current field
due to the magnetic field. The exact forms of the Green function
$G_{L_2,W_2}(\vec r, \vec r')$ in the cases $W_2 \ll L_2$ and $W_2 \gg
L_2$ are given in Eq.~(\ref{eq:Gsol}).

For a general source term we find:
$$
V_{(1)}^{(2)} (\vec r) = \int d^2 r' S(\vec r') G_{L_2,W_2}(\vec r, \vec r').
$$
By assumption, the source term in Eq.~(\ref{eq:V1}) is concentrated
near the origin and is local, we are interested in the voltage at
distances much larger than $a$ or $b$ and the thermal length $L_T$. In
addition, due to the boundary condition and conservation of charge
$\int d^2 r' S(\vec r') =0$, so we can use the a dipole approximation
to find:
$$
V_{(1)}^{(2)} (\vec r) =  \vec P \cdot \nabla' G (\vec r), \;\;\;
\vec P= \int d^2 r \vec r S (\vec r),
$$
\begin{equation}
\label{eq:V1a}
\nabla' G(\vec r) = \left. \nabla' G_{L_2,W_2}(\vec r, \vec r')
\right|_{\vec r'=0}.
\end{equation}

If we assume that the pinholes distribute uniformly over a rectangle
of size $a \times b$, the dipole moment associated with $g^t$ is given
by:
$$
{\vec P}_{\rm g} = \frac{J}{R_\perp  a b} \int_{-a/2}^{a/2} dx
\int_{-b/2}^{b/2} dy \left( x \; \hat x + y \; \hat y \right)
\left( R_\square x + R_H y\right) \Rightarrow
$$
\begin{equation}
\label{eq:Pg}
{\vec P}_{\rm g} = \frac{ J}{12 R_\perp} \left( a^2 R_\square \hat x
+ b^2 R_H \hat y \right)
\end{equation}
The dipole moment arising from the quantum corrections can be larger
than the leakage contribution since at low temperatures $L_T$ can be
larger than $a$ and than $b$. Using (\ref{eq:contenb}) and the
solution (\ref{sol:V0D}) for $V_0$ we arrive at:

\begin{eqnarray}
\label{eq:PQ}
{\vec P}_{\sf Q} = \int dx dy \times \nonumber \\
\sum_l\Bigl[ x{\sf Q}^l_{xx} \nabla_x \left( S^l_T(\vec r )
\nabla_x V_0^{(1)} \right)  \;\; \hat x + \nonumber \\
 y {\sf Q}^l_{xx} \nabla_y \left( S^l_T(\vec r
)\nabla_x V_0^{(1)} \right)  \;\; \hat y \Bigr] = \nonumber\\
 - {\sf Q} J L_T^2
\left(R_\square \hat x + R_H \hat y \right).
\end{eqnarray}
where in the last equality we have used integration by parts and the
normalization condition $\int d^2 r S^l_T(|\vec r|)=L_T^2 $.

\begin{figure}[h]
\vglue 0cm \hspace{0.05\hsize} \epsfxsize=1\hsize
\epsffile{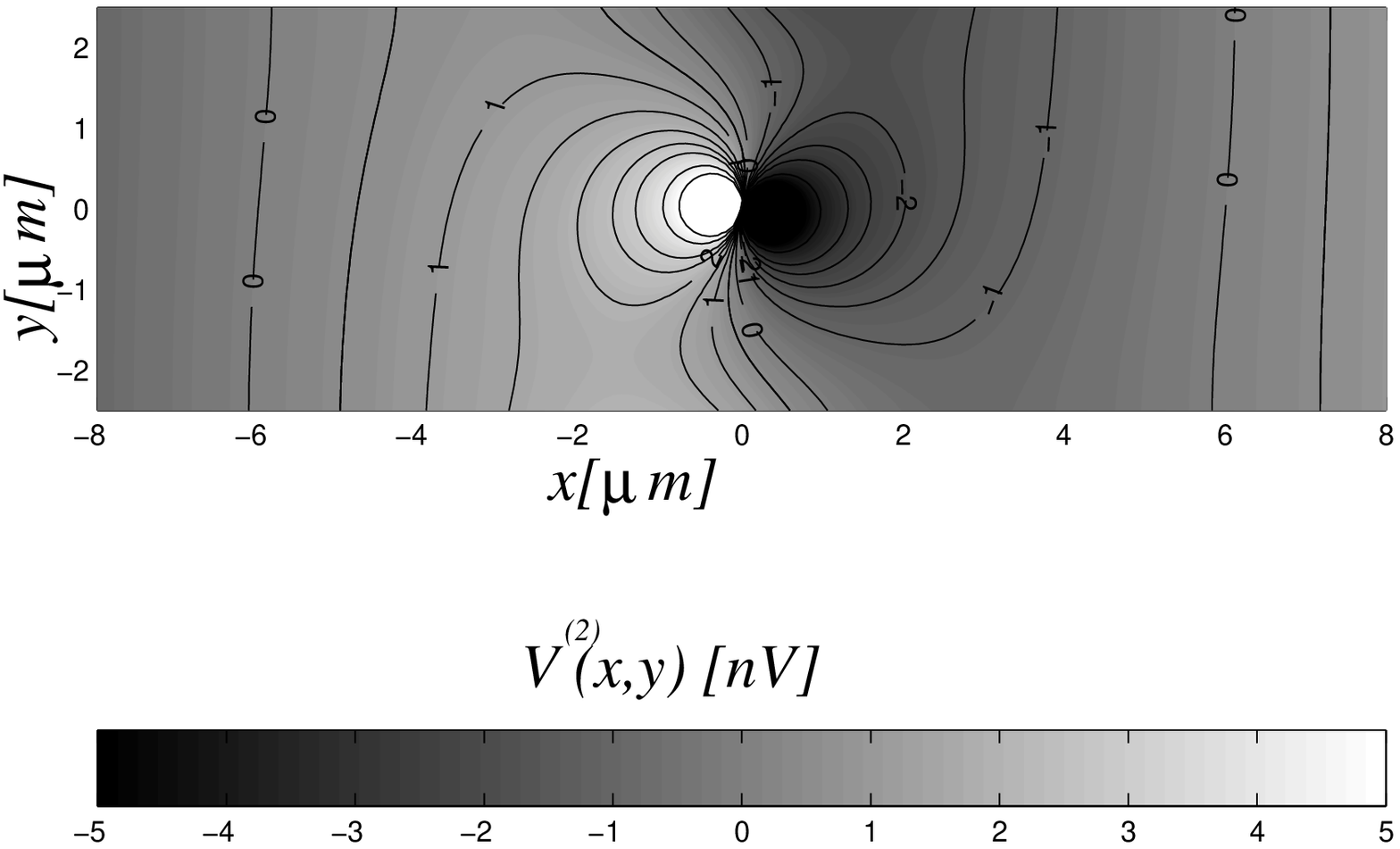} 
\refstepcounter{figure} \label{fg:v2con} \\
 {\small FIG.\ \ref{fg:v2con} A contour plot $V^{(2)}(x,y)$ according to
Eq.~(\ref{sol:V2}) at $T=0.5 K^\circ$.  We have used the same sample
parameters as in the introduction: $W_2= 5 \mu m$, mobility $\mu= 5
\times 10^4 {cm^2}/{Vs}$, electron density $n=4 \times 10^{10}
cm^{-2}$, $R_\perp=20 k\Omega$, $a=0.1\mu m, b=0.1\mu m$, $R_\square
\cong 3 k\Omega$, $\kappa d \cong 3$, with total current $I=0.1 \mu A$
and magnetic filed $H=0.05 {\rm T}$.}
\end{figure}

Even in the dipole approximation, the exact form of $V^{(2)}(x,y)$
depends on $W_2$ and $L_2$ since the information about the magnetic
field enters through the boundary conditions. For the ``parallel strip''
configuration, i.e., $L_2 \gg W_2$ using Eq.~\ref{eq:Gstrip} we
finally find, (keeping only terms linear in $H$):
\begin{eqnarray}
\label{sol:V2}
V^{2}(x,y) = J \rho_D x  \;\;\;\;\;\;\;\;\;\;\;\;\; \nonumber \\
- \frac{J}{2W_2}\left( \frac{ a^2}{12
R_\perp}+|{\sf Q}| L_T^2 \right)  \times  \;\;\;\;\;\;\;\;\;\;\;\;\;  \nonumber \\
\frac{\sinh(2 \pi x /W_2)}{ \cosh(2
\pi x /W_2)- \cos(2 \pi y /W_2)} R_\square^2 \nonumber \\
- \frac{J}{W_2} \left( \frac{ a^2+b^2}{12 R_\perp}    + 2|{\sf Q}| L_T^2
\right) \times  \;\;\;\;\;\;\;\;\;\;\;\;\;  \nonumber \\
 \frac{ \cosh(\pi x/W_2) \sin(\pi y /W_2)}{ \cosh(2 \pi x /W_2)-
\cos(2 \pi y /W_2)} R_\square R_H \nonumber \\
 +\frac{J}{2W_2}\frac{ a^2}{12 R_\perp}
\frac{ \sin(2 \pi y /W_2)}{ \cosh(2 \pi x /W_2)- \cos(2 \pi y /W_2)}
R_\square R_H
\end{eqnarray}
A contour plot of $V^{(2)}(x,y)$ with typical parameters is depicted in
Fig.~\ref{fg:v2con}. 
Since the potentials depend both on $x$ and $y$, the transresistances
depend on the locations where the potentials are measured.  Therefore
we have defined the integrated transresistances in Eqs.~(\ref{eq:Rt})
and (\ref{eq:RtH}).  In the ``parallel strip'' configuration far away
from the tunneling points, e.g., at $x \rightarrow \pm L_2/2$, the
potential practically does not depend on $y$. Thus, in the ``parallel
strip'' configuration while measuring $R_t$ the integration on the
potential is unnecessary.

In the ``cross'' configuration where $L_2 \ll W_2$, we use
Eq.~(\ref{eq:GL}) with $W_2 \rightarrow L_2$, now $x$ and $y$ change
their role. Far away from the tunneling points, e.g., at $y
\rightarrow \pm W_2/2$ the potential practically does not depend on
$x$.  In that ``cross'' geometry integration on the potential on the
boundary is not needed while measuring $R_t^H$.

\subsection{Tunneling through uniformly distributed  pinholes}

\label{se:uni}

Now we consider the case where the pinholes are distributed uniformly
in the sample. In this situation, we do not restrict ourselves to
first order in the tunneling conductance $R_\perp^{-1}$.  However, we
assume that $W_1=W_2=W,\; L_1=L_2=L$. In that case we can approximate
$\sum_l \left( g^l \cdots \right) \approx \int dx dy \left( 1/ R_\perp
  L W \cdots \right)$, and substitute the expression due to quantum
correction in Eq.~(\ref{eq:J1}) by:
\begin{equation}
\label{eq:sQ}
\sigma^Q_{ij} \equiv
\sum_l {\sf Q}^l_{ij} S_T^l(r)= {\sf Q}_{ij} \frac{L_T^2}{ L W} 
\end{equation}
in that case Eq.~(\ref{eq:ohmslawf}) should be read:
\begin{equation}
\label{eq:Ju}
{J}_i^\alpha ({\vec r}) = \left[\sigma_{ij} \delta_{\alpha \beta} +
\left( \sigma_{ij}^D + \sigma_{ij}^Q \right)X^{\alpha \beta} \right]
\bbox{\nabla}_j V_0^\beta ({\vec r})
\end{equation}
Now it is convenient to introduce upper and lower case variables:
\begin{eqnarray}
\label{eq:var}
 J_i = J^{(1)}_i + J^{(2)}_i, &\;\;\;\; &j_i =J^{(1)}_i - J^{(2)}_i
 \nonumber\\ \hat S = \hat \sigma + (\hat \sigma^D+\hat \sigma^Q),
 &\;\;\;\;& \hat s= \hat \sigma-(\hat \sigma^D+\hat \sigma^Q), \\
 V= V^{(1)} + V^{(2)},&\;\;& v= V^{(1)} - V^{(2)} \nonumber
\end{eqnarray}
Using now Eqs.~(\ref{eq:con}), (\ref{eq:Ju}), boundary conditions
(\ref{eq:bc12}), definitions (\ref{eq:var}) and the antisymmetry of
$\hat S$ and $ \hat s$ we arrive at the boundary problem:
\begin{eqnarray}
\label{EQ:V}
& &\;\; \Delta V = 0 \nonumber \\
\mbox { at } x=& \pm L/2 : &\;\; -S_{xx} \bbox{\nabla}_x V - S_{xy} \bbox{\nabla}_y V =J \\
\mbox { at } y=& \pm W /2: &\;\; -S_{yx} \bbox{\nabla}_x V - S_{yy} \bbox{\nabla}_y V =0 \nonumber
\end{eqnarray}
\begin{eqnarray}
\label{eq:v}
&&\Delta v = q^2v \nonumber \\ \mbox { at } x=& \pm L/2 :& -s_{xx}
\bbox{\nabla}_x v - s_{xy} \bbox{\nabla}_y v =J \\ \mbox { at } y=& \pm W/2 :&
-s_{yx} \bbox{\nabla}_x v - s_{yy} \bbox{\nabla}_y v = 0 \nonumber
\end{eqnarray}

where the characteristic momentum $q^2=2/(R_\perp L W s_{xx})$.
Eq.~(\ref{eq:v}) is a partial differential equation with boundary
conditions that are not Dirichlet nor Neumann type. A simple straight
forward solution is not available for general values of the magnetic
field, indicating the complexity of the current flow in layer 2 in the
presence of a magnetic field and uniform tunneling.

The detailed solution of Eqs.~(\ref{eq:v}) and (\ref{EQ:V}) (for weak
magnetic field) are given in App. \ref{app:Vv}. We find, then,
\begin{eqnarray}
\label{sol:V2uni}
V^{(2)}(x,y)= -\frac{J}{2} \left\{ \frac{1}{S_{xx}} x +
\frac{S_{xy}}{S_{xx}^2} y + \frac{1}{s_{xx} q} \times \right. \;\;\;\;\;\;\;\;\;\;\;\;\;\;\; \nonumber \\
\left.  \left[ \frac{\sinh(q
x)}{\cosh(q L/2)}+\frac{s_{xy}}{s_{xx}} \mbox{ thc } ( qL ) g
\left(\frac{x}{L}, \frac{y}{W}; q L, q W\right) \right] \right\}, \!\!\!
\end{eqnarray}
where $\mbox{thc} (x)$ and $g(u,z;a,b)$ are defined in
Eq.~({\ref{def:thc}) and(\ref{def:g}) respectively in
App.~\ref{app:Vv}.

\section{Conclusions}

To summarize, we have studied the effect of local pinholes on the drag
coefficient, $R_t$, and Hall drag coefficient, $R_t^H$, in bi--layer
dirty 2D systems at weak or vanishing magnetic field. We found that
there are three contributions to $R_t$: the classical drag, the
leakage contribution and the quantum contribution. The last two exist
only when there is tunneling through local pinholes between the
layers. At low temperature the classical drag vanishes. If $a$, the
characteristic size of the region where pinholes exist, is smaller
then the thermal length, $L_T$, but $L_T$ is small compared to the
overall system size, then the quantum contribution $\propto 1/T$ is
dominant.

The measurement of voltage in layer 2 in a typical drag experiment
does not allow current to flow perpendicular its edges. Therefore, the
classical drag does not lead to any Hall drag coefficient at low
temperatures. In contrast, in the presence of pinholes, we find
contributions to $R_t^H$ due to the leakage and the quantum
contributions. As for $R_t$, if $a< L_T$ the quantum contribution
$\propto 1/T$ is dominant at low temperatures.

The ``topographic map'' of the actual voltage in layer 2 is rather
complicated. We suggest therefore to study $R_t$ in a ``parallel
strip'' geometry (where the the width of layer 1 and layer 2 is
smaller than their length) and to study $R_t^H$ in a ``cross''
geometry where the width of layer 2 is larger than it length. In these
geometries, the voltage measurements are made far from the tunneling
region and the measured potentials do not depend on the precise
position of the probe along the boundary.

The drag and Hall drag measurements give direct information on the
quantum corrections arising from the interplay between disorder
interaction and tunneling.

In this work, we have studied the quantum contribution up to first
order in the measuring current, the tunneling conductance
$R_\perp^{-1}$, the magnetic field, and the interlayer interaction.
We have also estimated the limits for the validity of the
calculations.

  \acknowledgments
 
  It is our pleasure to thank A.~Kamenev, for useful discussions. YO
  is thankful for the support by the Rothschild fund. The work was
  supported by the NSF under grants no.\ DMR 94-16910, DMR 96-30064,
  DMR 97-14725, and DMR 98-09363.

\appendix
\section{Derivation of \bbox{\lowercase{$\uppercase{G_{L,W}}(\vec r, \vec r')$}} }
\label{app:G}
In this appendix we find the Green's function of the Poisson equation
in 2D:
\begin{equation}
\label{eq:G}
\sigma \Delta G_{L,W}(\vec r, \vec r') = \delta(\vec r - \vec r')
\end{equation}
with the boundary conditions
\begin{eqnarray}
\label{eq:bcG}
\sigma \nabla_x G_{L,W}(\pm L/2,y; \vec r') + \;\;\;\;\;\;\;\;\;\;
\nonumber \\
 \sigma_H \nabla_y G_{L,W}(\pm L/2,y; \vec r')= 0 \nonumber \\
 -\sigma_H \nabla_x G_{L,W}(x,\pm W/2; \vec r') + \;\;\;\;\;\;\;\;\;\;
  \nonumber \\
\sigma \nabla_y G_{L,W}(x,\pm W/2; \vec r')= 0
\end{eqnarray}
Without the boundary conditions (\ref{eq:bcG}) $G_{L, W}(\vec r, \vec
r')= 1/2 \pi \sigma \log |\vec r - \vec r'|$.  In the absence of a
magnetic field the boundary condition can be easily satisfied, by
introducing a series of image charges. Since the boundary conditions
are of Neumann type all the charges have the same sign.  In the
presence of magnetic field we should add to the image charges an image
circulation that cancels the circulation of the current field due to
the magnetic field. If we imagine a source charge situated on a black
square of an infinite chess board then on the black squares we have
charges of magnitude equal to the source charge and on the white
squares we should put charges and circulation sources. Formally, after
including the circulation sources we find:
\begin{eqnarray}
\label{eq:Gsol}
G_{L,W} (x,y; x',y')= \;\;\;\;\;\;\;\;\;\;\;\;\;\;\; \nonumber \\
\frac{1}{4 \pi \sigma}\sum_{n+m=e}\log \left[ (x- x'_n )^2 +(y-
y'_m)^2 \right] \ \nonumber \\ + \frac{1}{4 \pi \sigma} \left
[ \sum_{m+n =o} \frac{\sigma^2-\sigma_H^2}{\sigma^2+\sigma_H^2} \log
\left[(x- x'_n )^2 +(y- y'_m)^2 \right] \right. \nonumber \\ \left.  +
\frac{4 \sigma \sigma_H}{\sigma^2+\sigma_H^2}
\arctan[\frac{y-y'_m}{x-x'_n}] \right] \nonumber \\ x'_n=n L + x'
(-1)^n, \; y'_m= m W+ y' (-1)^m.
\end{eqnarray}
The summation in the first term is over even $m+n$ and in the second
over odd $m+n$.

  In the limiting case when $L \gg W $ we can take only the terms with
$n=0$. The summation of all sources can be performed analytically and
yields:

\begin{eqnarray}
\label{eq:Gstrip}
G_W(a,b; x,y) \equiv G_{L \rightarrow \infty,W}(a,b; x,y)= \nonumber
\\ \frac{1}{ 4 \pi \sigma}\Bigl\{ \log \left[\cosh\pi \frac{x-a}{ W}
-\cos \pi \frac{y-b}{W} \right] \nonumber \\ +
\frac{\sigma^2-\sigma_H^2}{\sigma^2+\sigma_H^2} \log \left[\cosh \pi
\frac{x-a}{W} +\cos\pi\frac{y+b}{W} \right] \nonumber \\ + \frac{4
\sigma \sigma_H}{\sigma^2+\sigma_H^2} \arctan\left[ \tan \pi
\frac{y+b}{2 W} \tanh \pi \frac{x-a}{ 2 W} \right] \Bigr\} \nonumber
\\ + {\rm const}
\end{eqnarray}
The constant is arbitrary since the boundary condition are of Neumann
type.  It can be directly checked that this Green function obeys the
boundary conditions (\ref{eq:bcG}) and the only singular point inside
the strip is at $x'$, $y'$. Close to it we find the usual logarithm
which is a solution of the Poisson equation with a delta function
source term.

  In the ``cross'' configuration when $W \gg L$ we have:
\begin{eqnarray}
\label{eq:GL}
G_{L} (x,y; x', y')= \;\;\;\;\;\;\;\;\;\;\; \nonumber \\ G_{W
\rightarrow L}( x \rightarrow -y, y \rightarrow x; x' \rightarrow -y',
y' \rightarrow x').
\end{eqnarray}

\section{Solution of Eqs. (\ref{EQ:V}) and 
{\lowercase{(\ref{eq:v})}} for weak magnetic field}

\label{app:Vv}
The solution of Eq.~(\ref{EQ:V}) can be written straightforwardly as
\begin{equation}
\label{sol:V}
V(x,y)= -\frac{JS_{xx}}{det} x- \frac{JS_{xy}}{det} y
\end{equation}
where $det= \det \hat S = S_{xx}^2+ S_{xy}^2$. This resembles the
solution of the usual single layer Hall problem.

At zero magnetic field the solution of Eq.~(\ref{eq:v}) is:
\begin{equation}
\label{eq:v0sol}
v_0(x,y)=-\frac{J}{s_{xx} q} \frac{\sinh(q x)}{\cosh(q L/2)}.
\end{equation}
For weak magnetic fields, expanding $v$ around $v_0$,
\begin{equation}
\label{def:v1}
v=v_0+\frac{s_{xy}}{s_{xx}} v_1
\end{equation}
and substituting in Eq.~(\ref{eq:v}) we find
\begin{eqnarray}
\label{eq:v1}
&\Delta v_1 = q^2 v_1& \nonumber \\ \mbox { at }& x= \pm L/2 :&
\bbox{\nabla}_x v_1 = 0 \\ \mbox { at }& y= \pm W/2 :& \bbox{\nabla}_y
v_1 = \bbox{\nabla}_x v_0, \nonumber
\end{eqnarray}
where we have kept only terms linear in the magnetic field $H$.
(Notice that the off diagonal elements of the matrices $\hat S$ and
$\hat s$ are proportional to $H$.)  Thus in the case of a weak
magnetic field the problem is reduced to a Neumann type boundary
problem, which we can solve by Fourier expansion in $x/L$. We find
\begin{equation}
\label{sol:v1}
v_1(x,y)=-\frac{J}{s_{xx} q } \mbox{thc} ( q L) g \left(\frac{x}{L},
\frac{y}{W}; q L, q W\right),
\end{equation}
where
\begin{equation}
\label{def:thc}
\mbox{thc}(x)=\tanh(x/2)/(x/2)
\end{equation}
\begin{eqnarray}
\label{def:g}
g(u,z;a,b)=&\frac{\sinh b z}{\cosh b/2}+
\;\;\;\;\;\;\;\;\;\;\;\;\;\;\;\;\;\;\;\;\;\;\;\;\;\;\;\; \nonumber \\
& 2 \sum_1^\infty a_n^3 \cos[ \pi n ( u-1)] \frac{\sinh a_n b z}{\cosh
a_n b/2}
\end{eqnarray}
and $a_n^2= a^2/(a^2+ \pi^2 n^2) $.  That gives:
\begin{eqnarray}
\label{sol:v}
v(x,y)&=&-\frac{J}{s_{xx} q}\left[ \frac{\sinh(q x)}{\cosh(q L/2)}+
\right. \nonumber \\ &&\left. \frac{s_{xy}}{s_{xx}} \mbox{ thc } ( qL
) g \left(\frac{x}{L}, \frac{y}{W}; q L, q W\right) \right]
\end{eqnarray}
Expressions (\ref{sol:V}) and (\ref{sol:v}) are the solution of the
boundary value problems (\ref{EQ:V}) and (\ref{eq:v}) in the limit of
a weak external magnetic field.

%\bibliography{/home/cmtv/oreg/ref/library}

\begin{thebibliography}{10}
\bibitem{DR:foot5}
We assume that the diameter of each pinhole is small compared to the electron
  inverse Fermi wave vector, $k_F^{-1}$, so that tunneling takes place through
  a single $s$-like channel. Thus, there can be no correlation between the
  initial and final momenta of a non--interacting electron tunneling through
  the pinhole. We also neglect any possible coherent interference due to
  tunneling through different pinholes that may lead to correlation between the
  initial and final momenta. This assumption should be valid at least in the
  case where the separation between the pinholes is larger than the mean free
  path in each layer. Coherent tunneling through different pinholes would also
  be suppressed by any differences in the Fermi momenta of the two layers due
  to asymmetry in the sample. It can also be suppressed deliberately by
  applying a magnetic field parallel to the
  layers\protect{\cite{DR:Boebinger91,DR:Berk94a}}. Finally we note that in the
  case where the pinholes are not small on the scale of the Fermi wavelength,
  or if there is momentum transfer due to coherent tunneling between different
  pinholes, we expect that the relative correction to the transconductance will
  be at most weakly dependent on the temperature for the case of greatest
  interest to us, where the region of pinholes has a diameter $a$ which is
  small compared to the thermal length $L_T=(D/T)^{1/2}$.
\end{thebibliography}

\end{document}